\documentclass[12pt,a4wide]{article}
\usepackage[usenames,dvipsnames]{pstricks} 
\usepackage{afterpage}
\usepackage{epsfig}
\usepackage{pst-grad} 
\usepackage{pst-plot} 
\usepackage[a4paper]{geometry}
\usepackage{float}
\usepackage[stable]{footmisc}
\usepackage[utf8x]{inputenc}
\usepackage{a4wide}
\usepackage{seqsplit}
\usepackage{amsmath,amstext,amsthm,amsfonts,amssymb}
\usepackage{mathrsfs}  
\usepackage{framed}
\usepackage{graphicx}
\usepackage{bbold}
\usepackage{bbm}
\usepackage{slashed} 
\usepackage{tensor} 
\usepackage{braket} 
\usepackage{amstext}
\usepackage{array}
\usepackage{graphicx}
\usepackage{setspace}
\usepackage{hyperref}
\usepackage{overpic}
\usepackage{bbm}
\usepackage{cite}
\usepackage{color}
\usepackage{lipsum} 
\usepackage{mdframed}
\usepackage{blkarray}
\usepackage{mathtools}
\usepackage{tikz}
\usepackage{url} 
\usepackage[normalem]{ulem}

\DeclareMathAlphabet{\mathbb}{U}{msb}{m}{n} 
\usepackage[linesnumbered,ruled,vlined]{algorithm2e}

\newcommand{\del}[0]{\partial}

\let\baraccent=\=
\renewcommand{\=}[1]{\stackrel{#1}{=}}
\newcommand{\id}[0]{\mathbbm{1}}

\usepackage{diagbox}
\newcommand\mathdiagbox[3][]{\hbox{\tabcolsep=\arraycolsep\diagbox[#1]{$#2$}{$#3$}}}

\makeatletter
\newcommand*\wt[1]{\mathpalette\wthelper{#1}}
\newcommand*\wthelper[2]{%
        \hbox{\dimen@\accentfontxheight#1%
                \accentfontxheight#11.1\dimen@
                $\m@th#1\widetilde{#2}$%
                \accentfontxheight#1\dimen@
        }%
}

\newcommand*\accentfontxheight[1]{%
        \fontdimen5\ifx#1\displaystyle
                \textfont
        \else\ifx#1\textstyle
                \textfont
        \else\ifx#1\scriptstyle
                \scriptfont
        \else
                \scriptscriptfont
        \fi\fi\fi3
}
\makeatother

\begin{document}
\pagestyle{plain}

	\makeatletter
	\@addtoreset{equation}{section}
	\makeatother
	\renewcommand{\theequation}{\thesection.\arabic{equation}}
	\pagestyle{empty}
	{\hfill \small MIT-CTP/5528}
	\vspace{0.5cm}
	
	\begin{center}
		
  {\LARGE \bf{Computational Mirror Symmetry} \\[10mm]}
	\end{center}

	\begin{center}
		\scalebox{0.95}[0.95]{{\fontsize{14}{30}\selectfont Mehmet Demirtas,$^{a,b}$ Manki Kim,$^{c}$ Liam McAllister,$^{d}$}} \vspace{0.35cm}
		\scalebox{0.95}[0.95]{{\fontsize{14}{30}\selectfont Jakob Moritz,$^{d}$ and Andres Rios-Tascon$^{d}$}}
	\end{center}

	\begin{center}
		\vspace{0.25 cm}

        \textsl{$^{a}$The NSF AI Institute for Artificial Intelligence and Fundamental Interactions} \\
        \textsl{$^{b}$Department of Physics, Northeastern University, Boston, MA 02115, USA} \\
        \textsl{$^{c}$Center for Theoretical Physics, MIT, Cambridge, MA 02139, USA}\\		\textsl{$^{d}$Department of Physics, Cornell University, Ithaca, NY 14853, USA}\\

		\vspace{1cm}
		\normalsize{\bf Abstract} \\[8mm]

	\end{center}

We present an efficient algorithm for computing the prepotential in compactifications of type II string theory on mirror pairs of Calabi-Yau threefolds in toric varieties.  Applying this method, we exhibit
the first systematic computation of genus-zero Gopakumar-Vafa invariants in compact threefolds with many moduli, including examples with up to 491 vector multiplets.

	\begin{center}
		\begin{minipage}[h]{15.0cm}

		\end{minipage}
	\end{center}
	\newpage
	\setcounter{page}{1}
	\pagestyle{plain}
	\renewcommand{\thefootnote}{\arabic{footnote}}
	\setcounter{footnote}{0}
	%
	%
	\tableofcontents
	\newpage

\section{Introduction}
 
Mirror symmetry is a profound fact about Calabi-Yau geometry that underlies much of what is known about string compactifications \cite{Candelas:1990rm,Witten:1991zz,Candelas:1993dm,HKTY,Candelas:1994hw,HKTY2,Strominger:1996it,Hori:2000kt,Hori:2003ic}.  Through mirror symmetry, one can compute certain quantum effects by performing classical period integrals: worldsheet instanton corrections to the prepotential in a compactification of type IIA string theory on a Calabi-Yau threefold $X$ can be determined from the periods of the $(3,0)$ form on the mirror 
$\wt{X}$.

The power of mirror symmetry has been brought to bear in understanding noncompact models, as well as compact models with very few moduli.  However, for a compact Calabi-Yau threefold with more than a handful of moduli in vector multiplets, carrying out the mirror map with pen and paper is not feasible.  Indeed, even with a computer this general case has been inaccessible. 
This state of affairs has limited the use of mirror symmetry in understanding the landscape of string compactifications.

The purpose of this work is to provide a practical computational algorithm that implements the mirror map in a very large class of Calabi-Yau threefolds, order by order in an expansion around large volume/large complex structure.  We will study mirror pairs of compact Calabi-Yau threefold hypersurfaces in toric varieties, following Batyrev \cite{Batyrev:1993oya}, and will compute the prepotential by building on ideas of Hosono, Klemm, Theisen, and Yau (HKTY) \cite{HKTY,HKTY2}. Our algorithm remains practical even for the hypersurface with the largest known number of moduli.

There are three principal obstacles to computing the data of a mirror pair in this setting, when $h^{1,1}(X)=h^{2,1}(\wt{X})$ is large.  First, one needs to compute the intersection numbers of $X$, which by assumption has many K\"ahler moduli.
Second, the procedure of \cite{HKTY}, which involves computing a fundamental period and using properties of the Picard-Fuchs system, requires that the Mori cone of $X$ be simplicial, a condition that is almost never met for $h^{1,1}(X) \gg 1$.  Third, carrying out the mirror map and computing enumerative invariants of $X$ requires examining a number of lattice sites in the Mori cone of $X$ that is exponential in $h^{1,1}(X)$.

The first obstacle, obtaining the intersection numbers of a threefold hypersurface with many K\"ahler moduli, was first solved through the software package \texttt{CYTools} \cite{Demirtas:2022hqf}.  In the present work, we overcome the two remaining obstacles: we generalize the approach of \cite{HKTY} to threefolds with non-simplicial Mori cones, and we present an efficient algorithm for carrying out the mirror map in this setting.  Our method applies to any Calabi-Yau threefold hypersurface, and remains practical on a laptop even in cases
with the largest-known number of complex structure moduli, i.e.~$h^{2,1}(\wt{X})=491$.

The organization of this paper is as follows.  
In \S\ref{sec:prepotential} we recall properties of the prepotential in type II compactifications on Calabi-Yau threefolds. 
In \S\ref{sec:batyrev} we set notation for discussing Calabi-Yau hypersurfaces in toric varieties.
In \S\ref{sec:hktyandbeyond} we recall the method introduced by HKTY \cite{HKTY} for computing the prepotential via mirror symmetry, and we explain a simplifying assumption made in \cite{HKTY} that limits the scope of this method.  We then explain how to generalize the idea of \cite{HKTY} to an arbitrary Calabi-Yau threefold hypersurface.
In \S\ref{sec:algorithm} we present a computationally efficient algorithm for computing the prepotential.
For certain curves whose toric description is simple, one can determine the Gopakumar-Vafa invariants directly, without the procedure given in \S\ref{sec:hktyandbeyond}.  We describe this complementary approach in \S\ref{sec:toric}.
In \S\ref{sec:examples} we illustrate our method in a few examples.  
We conclude in \S\ref{sec:conclusions}.
Appendix \S\ref{sec:Appendix_cformulas} contains useful formulas for certain coefficients in the expansion of the periods.

\section{The Prepotential in Type II Compactifications}\label{sec:prepotential}

In this section we set notation and review properties of type II compactifications on Calabi-Yau 
threefolds, focusing on the prepotential for the vector multiplet sector.  We recall the definition of Gopakumar-Vafa (GV) invariants and collect basic facts about mirror symmetry.

\subsection{Calabi-Yau compactifications of type IIA string theory}

\subsubsection{Setup}

We  consider type IIA string theory on a Calabi-Yau threefold $X$.
Denoting by $h^{p,q}$ the Hodge numbers of $X$,
we take $\bigl\{[\mathcal{C}^a]\bigr\}_{a=1}^{h^{1,1}}$ to be a basis of $H_2(X,\mathbb{Z})$, and we write $\bigl\{[H_a]\bigr\}_{a=1}^{h^{1,1}}$ for the dual basis of $H^2(X,\mathbb{Z})$:
\begin{equation}\label{eq:twocycbasis}
    \int_X [H_a]\wedge [\mathcal{C}^b]={\delta_a}^b\, .
\end{equation}
The moduli space of K\"ahler structures on $X$ 
has dimension $h^{1,1}$
and is parameterized by the K\"ahler form $J$, which takes values in the K\"ahler cone $\mathcal{K}_X\subset H^{1,1}(X,\mathbb{R})$. We denote by $B_2$ the ten-dimensional Kalb-Ramond two-form of type IIA string theory, and
we define the complexified K\"ahler form $J_c:=B_2+iJ$.

The low energy effective field theory preserves eight supercharges, and contains the supergravity multiplet $\mathcal{W}$, as well as $h^{1,1}$ vector multiplets $\mathcal{V}^a$, $a=1,\ldots,h^{1,1}$, and $h^{2,1}+1$ hypermultiplets $\mathcal{H}^i$, $i=0,\ldots,h^{2,1}$, one of which contains the string coupling. The gauge group at a generic point in moduli space is $U(1)^{h^{1,1}+1}$, with $h^{1,1}$ gauge fields living in vector multiplets, and one further gauge field in the gravity multiplet. In this paper we will focus on the gravity and vector multiplet sector.

The massless $U(1)$ gauge fields $(A^0,A^a)$ come from the dimensional reduction of the ten-dimensional $p$-form potentials 
\begin{equation}\label{eq:C1C3_KKreduction}
	C_1\rightarrow A^0(x)\, ,\quad C_3\rightarrow \sum_{a=1}^{h^{1,1}}A^a(x)\wedge [H_a]\, ,
\end{equation}
and their electric-magnetic duals $(A_0,A_a)$ can likewise be thought of as the dimensional reduction of the dual $p$-form potentials $C_7$ and $C_5$,
\begin{equation}\label{eq:C5C7_KKreduction}
	C_7\rightarrow A_0(x)\wedge w_X\, ,\quad C_5\rightarrow \sum_{a=1}^{h^{1,1}}A_a(x)\wedge [\mathcal{C}^a]\, ,
\end{equation}
where $w_X$ is the volume form of $X$ normalized such that $\int_X w_X=1$.

We may parameterize the vector multiplet moduli space by $h^{1,1}$ complex fields
\begin{equation}\label{eq:udef}
    u^a:=\int_{X}J_c\wedge [\mathcal{C}^a]\, ,
\end{equation}
defined modulo integers, $u^a\simeq u^a+1$. The imaginary parts $t^a:=\text{Im}(u^a)$ are coordinates on the K\"ahler cone $\mathcal{K}_X$, and the real parts parameterize the flat B-field.

Extended supersymmetry guarantees that the vector multiplet moduli space is projective special K\"ahler, and thus its K\"ahler metric derives from a K\"ahler potential $K(u,\bar{u})$ that can be written in terms of a holomorphic prepotential $\mathcal{F}(u)$ as
\begin{equation}\label{eq:kahler_potential}
    K=-\log\Bigl(2i(\mathcal{F}-it^a\mathcal{F}_a)+c.c.\Bigr)\, ,\quad \mathcal{F}_a:=\del_{u^a}\mathcal{F}\, .
\end{equation}
The string coupling lives in a hypermultiplet, and so the prepotential is tree-level exact in the string loop expansion.  However, as the K\"ahler moduli of $X$ live in vector multiplets, the prepotential $\mathcal{F}$ for the vector multiplets receives both perturbative and nonperturbative corrections in $\alpha'$:
\begin{equation}\label{eq:prepotential_split}
	\mathcal{F}= \mathcal{F}_{\text{pert.}} + \mathcal{F}_{\text{inst.}}\, .
\end{equation}
In addition to governing the metric on moduli space, the prepotential   also computes the central charges $\mathcal{Z}_{\Vec{Q}}$ of certain BPS states.  
We consider a bound state of D$p$-branes wrapped on $p$-cycles, with $p\in\{0,2,4,6\}$, potentially carrying worldvolume fluxes, and with electric and magnetic charges $\Vec{Q}$ (elements of K-theory).
Then we have
\begin{equation}
    \mathcal{Z}_{\Vec{Q}}=\Vec{Q}^T\cdot \Vec{\Pi}_{\text{IIA}}\, ,
\end{equation}
in terms of the \emph{period vector}
\begin{equation}\label{eq:iiaper}
    \Vec{\Pi}_{\text{IIA}}:=\begin{pmatrix}
        2\mathcal{F}-u^a \mathcal{F}_a\\
        \mathcal{F}_a\\
        1\\
        u^a
    \end{pmatrix}\, .
\end{equation}
One may write the K\"ahler potential \eqref{eq:kahler_potential} in terms of the period vector,
\begin{equation}
    K=-\log\left(i\Vec{\Pi}^\dagger_{\text{IIA}} \cdot \mathcal{I}^{-1} \cdot \Vec{\Pi}_{\text{IIA}} \right)\, ,
\end{equation}
where  
\begin{equation}\label{eq:symplectic_pairing}
    \mathcal{I}:=\begin{pmatrix}
    0 & \id \\
    -\id & 0
\end{pmatrix}
\end{equation}
is the symplectic electric-magnetic pairing. Given a pair of D-branes with electric-magnetic charge vectors $(\Vec{Q},\Vec{Q'})$, realized by wrapping a pair of cycles $(\Sigma,\Sigma')$ with worldvolume fluxes $(\mathcal{F},\mathcal{F}')$ this pairing is equal to the index of the fluxed Dirac operator on $\Sigma\cap \Sigma'$ \cite{Minasian:1997mm},
\begin{equation}\label{eq:Dirac_index_pairing}
    \Vec{Q}^T\cdot \mathcal{I}\cdot \Vec{Q}'= \int_{\Sigma\cap \Sigma'} e^{\mathcal{F}-\mathcal{F}'}\hat{A}\bigl(T(\Sigma\cap\Sigma')\bigr)\,,
\end{equation}
where \cite{Freed:1999vc}
\begin{equation}
\mathcal{F}-\frac{1}{2}c_1\bigl(T(\Sigma\cap\Sigma'\bigr) \in H^2(\Sigma\cap\Sigma',\mathbb{Z})\,.
\end{equation}

Finally, we note that the holomorphic prepotential is only the lowest component of an infinite tower of F-terms that contribute to the Wilsonian effective action \cite{Bershadsky:1993cx,Antoniadis_1995}
\begin{equation}
	S_{\text{eff}}\supset -i\int d^4x \int d^4\theta \sum_{g=0}^{\infty}\mathcal{F}_g\, \mathcal{W}^{2g}\, ,
\end{equation}
that involve only the gravity multiplet and the vector multiplets via holomorphic functions $\mathcal{F}_g\equiv \mathcal{F}_g(\mathcal{V}^a)$. These are given by the genus $g$ contributions to the free energy of the topological string \cite{Witten:1991zz}. At low energies, the most relevant term in the derivative expansion is the prepotential $\mathcal{F}_0\equiv \mathcal{F}$, which results from worldsheets of genus $g=0$.

\subsubsection{Perturbative type IIA prepotential}\label{sec:iiapert}
Next, we will explain how to compute the perturbative contributions to the type IIA prepotential \eqref{eq:prepotential_split}.

At tree level in $\alpha'$ the prepotential is simply
\begin{equation}\label{eq:ftree}
	\mathcal{F}_{\text{tree}}(u)=-\frac{1}{3!}\kappa_{abc}u^au^bu^c\, 
\end{equation}
in terms of the  triple intersection numbers $\kappa_{abc}:=\int_X [H_a]\wedge [H_b]\wedge [H_c]$.

The exact prepotential is strongly constrained by monodromies in K\"ahler moduli space, which amount to the Witten effect in the four-dimensional effective theory.\footnote{For a related discussion see \cite{Corvilain:2018lgw}.} In order to understand these monodromies, one first sets up an integer basis of electrically and magnetically charged particle states. These are furnished by D6-D4-D2-D0 bound states wrapped on even-dimensional cycles in $X$. 

Specifically, we may choose an integer basis 
\begin{equation}\label{eq:D_even_basis}
	\{m_0,m_a,e^0,e^a\}\equiv \{m_A,e^A\}
\end{equation}
of the $2h^{1,1}+2$-dimensional electric-magnetic charge lattice as follows. We take $m_0$ to be a D6-brane wrapped on $X$ without worldvolume fluxes; the $m_a$ to be D4-branes wrapped on representatives of our basis of four-cycle classes $[H_a]$, with Freed-Witten canceling worldvolume fluxes $\mathcal{F}_a=-\frac{1}{2}[H_a]|_{H_a}$ turned on; the $e^a$ to be D2-branes wrapped on representatives of the dual basis of curve classes $[\mathcal{C}]^a$; and $e^0$ to be a D0 brane. These basis elements undergo monodromies (i.e. the Witten effect \cite{Witten:1979ey}) at large volume that are governed by the anomalous Chern-Simons (CS) term of D-branes \cite{Green:1996dd,Cheung:1997az}:
\begin{equation}\label{eq:CS-action}
	S_{CS}^{\Sigma}=\pm \frac{2\pi}{\ell_s^{p+1}}\int_{\Sigma} \,\,\sum_{p\,\,\, \text{odd}} C_p\wedge e^{B_2+\mathcal{F}_{\Sigma}}\wedge \sqrt{\frac{\hat{A}(T\Sigma)}{\hat{A}(N\Sigma)}}\, ,
\end{equation}
where $\mathcal{F}_{\Sigma}$ is the worldvolume flux on the D-brane, $T\Sigma$ is the tangent bundle, $N\Sigma$ is the normal bundle, and $\hat{A}$ denotes the A-roof genus.
The monodromy action is as follows: under $B_2\rightarrow B_2+w$, with $w\in H^2(X,\mathbb{Z})$, the integrand picks up a factor of $e^w$. 

The CS action \eqref{eq:CS-action} is the de Rham inner product on $H^\bullet(X)$ between the vector of gauge fields --- expressed as the poly-form $\sum_p C_p$ written in terms of the gauge fields as in \eqref{eq:C1C3_KKreduction} and \eqref{eq:C5C7_KKreduction} --- and the BPS particle charge, expressed as the poly-form $e^{B_2+\mathcal{F}_{\Sigma}}\wedge \sqrt{\frac{\hat{A}(T\Sigma)}{\hat{A}(N\Sigma)}}$. Thus, we can understand the $B_2$ monodromy as a linear map acting on the charge lattice
\begin{equation}
    Q_{\Sigma}=e^{B_2+\mathcal{F}_{\Sigma}}\wedge \sqrt{\frac{\hat{A}(T\Sigma)}{\hat{A}(N\Sigma)}}\mapsto e^w Q_{\Sigma}\, .
\end{equation}
The central charge $\mathcal{Z}$ of a particle state must undergo the very same monodromies, and be holomorphic in the complex coordinates on vector multiplet moduli space $u^a=\int_X (B_2+iJ)\wedge [\mathcal{C}]^a$. One thus concludes that the central charge of a D-brane wrapped on a cycle $\Sigma$ is given by
\begin{equation}
	\mathcal{Z}_{\Sigma}=\int_{\Sigma} e^{B_2+iJ+\mathcal{F}_{\Sigma}}\wedge \sqrt{\frac{\hat{A}(T\Sigma)}{\hat{A}(N\Sigma)}}\, ,
\end{equation}
modulo terms that are single-valued under the monodromies.

The central charges of the elements of the basis \eqref{eq:D_even_basis} 
are
\begin{equation}
	\vec{\mathcal{Z}}= \begin{pmatrix}
		\frac{1}{3!}\kappa_{abc}u^au^bu^c+\frac{1}{24}u^a c_a\\
		-\frac{1}{2}\kappa_{abc}u^bu^c+\frac{1}{2}\kappa_{aab}u^b -\frac{1}{6}\kappa_{aaa}-\frac{1}{24}c_a\\
		1\\
		u^a
	\end{pmatrix}\, .
\end{equation}
However, the basis  \eqref{eq:D_even_basis} is not in general symplectic, in the sense that the electric-magnetic pairing deviates from the standard form \eqref{eq:symplectic_pairing}.
We will now determine an integer change of basis that brings the symplectic electric-magnetic pairing to the standard form.

In the basis \eqref{eq:D_even_basis}  the pairwise Dirac indices \eqref{eq:Dirac_index_pairing} evaluate to
\begin{equation}
	\tilde{\mathcal{I}}=\begin{pmatrix}
		\Lambda & \id\\
		-\id & 0
	\end{pmatrix}\, ,\quad \text{with} \quad \Lambda = \begin{pmatrix}
		0     & -i_b\\
		i_a  & h_{ab}
	\end{pmatrix}\, ,
\end{equation}
where
\begin{equation}
	h_{ab}=\frac{1}{2}(\kappa_{abb}-\kappa_{aab})\, ,\quad i_a=\frac{1}{12}c_a+\frac{1}{6}\kappa_{aaa}\, .
\end{equation}
It is then straightforward to find the integer change of basis that brings the symplectic pairing to canonical form. In this new basis the vector of central charges becomes
\begin{equation}
	\vec{\mathcal{Z}}'=\begin{pmatrix}
		\id & M\\
		0 & \id
	\end{pmatrix}\cdot \vec{\mathcal{Z}}\, ,\quad \text{with} \quad M=\begin{pmatrix}
		0 & 0 \\
		i_a & h_{ab}\Theta(b-a)\, .
	\end{pmatrix}
\end{equation}
Up to terms that do not affect the monodromies, one finds
\begin{equation}
	\vec{\mathcal{Z}}'=\begin{pmatrix}
		\frac{1}{3!}\kappa_{abc}u^au^bu^c+\frac{1}{24}c_au^a\\
		-\frac{1}{2}\kappa_{abc}u^bu^c+a_{ab}u^b+\frac{1}{24}c_{a}\\
		1\\
		u^a
	\end{pmatrix}\, ,
\end{equation}
with
\begin{equation}
	a_{ab}:= \begin{cases}
		\frac{1}{2}\kappa_{aab} & a\geq b\\
		\frac{1}{2}\kappa_{abb} & a<b
	\end{cases}\, .
\end{equation}
This central charge vector indeed derives from a prepotential,
\begin{equation}
	\hat{\mathcal{F}}(u)=-\frac{1}{3!}\kappa_{abc} u^a u^b u^c+\frac{1}{2}a_{ab}u^au^b+\frac{1}{24}u^ac_a\, .
\end{equation}
The most general ansatz for the vector of central charges that is compatible with the monodromies of $\vec{\mathcal{Z}}'$ is
\begin{equation}
	\vec{\Pi}_{\text{IIA}}=\vec{\mathcal{Z}}'+\begin{pmatrix}
		2 \delta \mathcal{F}-u^a \del_a\delta \mathcal{F}\\
		\del_a \delta \mathcal{F}\\
		0\\
		0
	\end{pmatrix}\, ,
\end{equation}
with $\delta \mathcal{F}(u)$ a holomorphic function in the $u^a$ that is invariant under the monodromies. Thus, the exact prepotential reads
\begin{equation}
	\mathcal{F}:=\hat{\mathcal{F}}+\delta \mathcal{F}\, ,
\end{equation}
with  
\begin{equation}\label{delf}
	\delta \mathcal{F}=const. +\sum_{0 \neq q\in H_2(X,\mathbb{Z})}A_q e^{2\pi i q\cdot u}\, .
\end{equation}
The requirement 
that $\mathcal{F}:=\hat{\mathcal{F}}+\delta \mathcal{F}$ should asymptote to the tree level result
\eqref{eq:ftree} for $\text{Im}(u) \gg 1$
implies that the sum in \eqref{delf} must only include points $q \in H_2(X,\mathbb{Z})$ for which $q\cdot \text{Im}(u) \ge 0$ for all $\text{Im}(u)$ in the K\"ahler cone.
In other words, the sum is restricted to the Mori cone of $X$.
We can identify $\delta \mathcal{F}$ as the sum of a perturbative $\mathcal{O}(\alpha'^3)$ correction, and a tower of worldsheet instantons.

The $\mathcal{O}(\alpha'^3)$ correction is in fact well known \cite{Font:1992uk,Klemm:1992tx,Becker:2002nn}, and thus to all orders in perturbation theory one has
\begin{equation}\label{eq:pert_prepotential}
	\mathcal{F}_{\text{pert.}}(u)=-\frac{1}{3!}\kappa_{abc}u^au^bu^c+\frac{1}{2}a_{ab}u^au^b+\frac{c_a}{24} u^a+\frac{\zeta(3)\chi(X)}{2(2\pi i)^3}\, ,
\end{equation}
with
\begin{equation}
	c_a:=\int_{\Sigma_{4}^a}c_2(X)\, ,\quad \text{and} \quad  \chi(X)=\int_X c_3(X)\, .
\end{equation}
Here $\zeta(z)$ denotes the Riemann zeta function. We will say more about the non-perturbative part $\mathcal{F}_{\text{inst.}}$ in \S\ref{sec:GV_WS_instantons}.

We pause to stress a crucial point: the large volume monodromies $B_2\rightarrow B_2+w$ with $w\in H^2(X,\mathbb{Z})$, together with the topological $\mathcal{O}(\alpha'^3)$ correction, uniquely\footnote{The result is unique modulo symplectic transformations.} determine the period vector of central charges $\Vec{\Pi}_{\text{IIA}}$ \eqref{eq:iiaper} expressed in an \emph{integral symplectic basis}, up to corrections that fall off exponentially in the large volume limit: we have seen that the perturbative part of the prepotential $\mathcal{F}_{\text{pert.}}(u)$ is given by \eqref{eq:pert_prepotential}.
The data that computes \eqref{eq:pert_prepotential}  is given by purely classical geometric data of $X$, in the form of its intersection numbers and Chern classes. 
 
\subsubsection{Instanton corrections to the type IIA prepotential}\label{sec:GV_WS_instantons}
The instanton contributions to the prepotential result from genus zero worldsheets wrapping curves in non-trivial classes $[\mathcal{C}]\in H_2(X,\mathbb{Z})$:
\begin{equation}\label{eq:finst}
	\mathcal{F}_{\text{inst.}}(u)=-\frac{1}{(2\pi i)^3}\sum_{[\mathcal{C}]\in \mathcal{M}_X}\text{GW}^{0}_{[\mathcal{C}]}\, q^{[\mathcal{C}]}\, ,
\end{equation}
where $\text{GW}^{0}_{[\mathcal{C}]} \in \mathbb{Q}$ is the genus zero \textit{Gromov-Witten} (GW) invariant of the curve class $[\mathcal{C}]$, $\mathcal{M}_X$ is the \textit{Mori cone} of effective curves, and
\begin{equation}
	q^{[\mathcal{C}]}:= \exp\left(2\pi i \int_{\mathcal{C}} J_c \right) = \exp\biggl(2\pi i \beta_{a}^{[\mathcal{C}]} u^a \biggr)\, ,
\end{equation}
where $[\mathcal{C}] = [\mathcal{C}^a] \beta_{a}^{[\mathcal{C}]}$ in terms of the basis 
$\bigl\{[\mathcal{C}^a]\bigr\}$ and coefficients $\beta_{a}^{[\mathcal{C}]}$, and the coordinates $u^a$ were defined in \eqref{eq:udef}.

Similarly, the higher-genus contributions $\mathcal{F}_g$ receive non-perturbative contributions from genus $g$ worldsheets wrapping effective curves, with associated higher genus GW invariants $\text{GW}^{g}_{[\mathcal{C}]}$.

As first observed in \cite{Candelas:1990rm}, the series in \eqref{eq:finst} can be re-expressed in terms of \emph{integer} invariants $\text{GV}^{0}_{[\mathcal{C}]}$,
\begin{equation}\label{eq:FinstIIA}
	\mathcal{F}_{\text{inst.}}(u)=-\frac{1}{(2\pi i)^3}\sum_{[\mathcal{C}]\in \mathcal{M}_X} \text{GV}^{0}_{[\mathcal{C}]} \, \text{Li}_3\left(q^{[\mathcal{C}]}\right)\, ,
\end{equation}
with polylogarithm $\text{Li}_k(q):=\sum_{l=1}^\infty q^l/l^k$. The $\text{GV}^{0}_{[\mathcal{C}]} $ are the genus zero \emph{Gopakumar-Vafa (GV) invariants}.

The fact that $\text{GV}^{0}_{[\mathcal{C}]} \in \mathbb{Z}$ is explained by the interpretation of the $\text{GV}^{0}_{[\mathcal{C}]}$ as certain BPS indices \cite{Gopakumar:1998ii,Gopakumar:1998jq}. As type IIA string theory on $X$ is dual to M-theory on $X\times S^1$, one can view the four-dimensional theory as a circle compactification of a five-dimensional $\mathcal{N}=1$ theory that arises from M-theory on $X$. The F-terms of the five-dimensional theory are classically exact, so one may view the non-perturbative contributions to the four-dimensional F-terms as particle instantons from BPS particles of the five-dimensional theory traveling around the compactification circle in Euclidean time. 

The BPS indices $\text{GV}^{g}_{[\mathcal{C}]}$ are defined as follows. Given a fixed effective curve class $[\mathcal{C}]$, quantization of the degrees of freedom on M2-branes wrapping curves in  $[\mathcal{C}]$ gives rise to a BPS particle spectrum that transforms in the massive little group $\text{Spin}(4)\simeq SU(2)_L\times SU(2)_R$ in a representation 
\begin{equation}
	\left(2(0)_L\oplus \left(\frac{1}{2}\right)_L\right)\otimes\sum_{j_L,j_R}N_{j_L,j_R}^{[\mathcal{C}]}(j_L)_L\otimes (j_R)_R\, ,
\end{equation}
with degeneracies $N_{j_L,j_R}^{[\mathcal{C}]}$. The contribution to the index is given by the decomposition
\cite{Gopakumar:1998jq}
\begin{equation}
	\sum_{j_L,j_R}(-1)^{2j_R}(2j_R+1)N_{j_L,j_R}^{[\mathcal{C}]}(j_L)_L=\sum_{g}\text{GV}^g_{[\mathcal{C}]} \left(2(0)_L+\left(\frac{1}{2}\right)_L\right)^g\, .
\end{equation}
The GV invariants and GW invariants are related by \cite{Gopakumar:1998ii,Gopakumar:1998jq}
\begin{equation}
	\sum_{g=0}^\infty \lambda^{2g-2} \sum_{[\mathcal{C}]\in \mathcal{M}_X}\text{GW}_{[\mathcal{C}]}^g\,q^{[\mathcal{C}]}=\sum_{g=0}^\infty \sum_{[\mathcal{C}]\in \mathcal{M}_X} \text{GV}_{[\mathcal{C}]}^g\sum_{k=1}^{\infty}\frac{1}{k}\left(2\sin\left(\frac{k\lambda}{2}\right)\right)^{2g-2} q^{k[\mathcal{C}]}\, .
\end{equation}

\subsection{Calabi-Yau compactifications of type IIB string theory}

Next, we consider compactifying type IIB string theory on the mirror Calabi-Yau, which we will denote by $\wt{X}$. We write $\Tilde{h}^{p,q}$ for the Hodge numbers of $\wt{X}$, which obey $\Tilde{h}^{2,1}=h^{1,1}$ and $\Tilde{h}^{1,1}=h^{2,1}$.  We denote the nowhere-vanishing holomorphic three-form on $\wt{X}$, which is unique up to overall scale, by $\wt{\Omega}$.

Mirror symmetry implies that the physical theory in type IIB compactification on $\wt{X}$ is equivalent to that in compactification of type IIA string theory on $X$, discussed hitherto.
However, the geometric interpretation of the low energy degrees of freedom is completely different: 
the vector multiplet moduli space of $\wt{X}$ is now the moduli space  space of complex structures. As both the type IIB string coupling and the K\"ahler moduli of $\wt{X}$ live in hypermultiplets, the vector multiplet moduli space is tree-level exact both in the string loop and the $\alpha'$ expansions, i.e.~it is completely classical.

We let $\bigl\{[\Tilde{\alpha}^A],[\Tilde{\beta}_A] \bigr\}_{A=0}^{\Tilde{h}^{2,1}}$ be a symplectic basis of $H^3(\wt{X},\mathbb{Z})$, i.e.
\begin{equation}
    \int_{\wt{X}} [\Tilde{\alpha}^A]\wedge [\Tilde{\beta}_B]={\delta^A}_B\, , \quad \int_{\wt{X}} [\Tilde{\alpha}^A]\wedge [\Tilde{\alpha}^B]=\int_{\wt{X}} [\Tilde{\beta}^A]\wedge [\Tilde{\beta}^B]=0\, ,
\end{equation}
and we define the periods of the holomorphic three-form
\begin{equation}\label{eq:iibperdef}
    \Vec{\Pi}_{\text{IIB}}:=\begin{pmatrix}
        \int_{\wt{X}}\wt{\Omega}\wedge \beta_A\\
        \int_{\wt{X}}\wt{\Omega}\wedge \alpha^A
    \end{pmatrix}\, .
\end{equation}
At generic points in moduli space we may use any subset of $\Tilde{h}^{2,1}$ of the periods as local coordinates on moduli space. 

Mirror symmetry states that for a suitably chosen basis, and for an appropriate normalization of $\wt{\Omega}$, we have
\begin{equation}\label{eq:mirrorsays}
    \Vec{\Pi}_{\text{IIA}}=\Vec{\Pi}_{\text{IIB}}\equiv \Vec{\Pi}\, .
\end{equation}
By definition, the large volume expansion of $\Vec{\Pi}_{\text{IIA}}$ is equivalent to the \emph{large complex structure} (LCS) expansion of $\Vec{\Pi}_{\text{IIB}}$. Indeed, according to the SYZ conjecture \cite{Strominger:1996it}, a Calabi-Yau $n$-fold is $T^n$-fibered at sufficiently large complex structure, and mirror symmetry can be understood as T-duality along all $n$ legs of the torus fiber, thus mapping type IIA and type IIB string compactifications into each other for $n$ odd.

The exact K\"ahler metric on moduli space is equal to the classical Weil-Petersson metric derived from
\begin{equation}\label{eq:kahler_potential_IIB}
    K=-\log\left(-i\int_{\wt{X}} \wt{\Omega}\wedge \overline{\wt{\Omega}}\right)\, .
\end{equation}
The mirror dual of the basis \eqref{eq:D_even_basis}  of electric-magnetically charged states in type IIA is represented by D3-branes wrapped on the basis of cycles $\bigl\{[\Tilde{\alpha}^A],[\Tilde{\beta}_A] \bigr\}_{A=0}^{\Tilde{h}^{2,1}}$, and the Dirac index becomes the classical geometric intersection product.

In general it is difficult to compute the periods of the holomorphic three-form in an integer symplectic basis. However, following \cite{HKTY}, we will see in \S\ref{sec:hktyandbeyond} that one can compute the periods in an arbitrary $\mathbb{C}$-basis systematically around LCS. Combining this with the perturbative type IIA result \eqref{eq:pert_prepotential} then allows one to compute the period vector systematically in an integral symplectic basis.

\section{Mirror Symmetry for Hypersurfaces}\label{sec:batyrev}
 
As we have reviewed in \S\ref{sec:prepotential}, one can use the fundamental mirror symmetry relation \eqref{eq:mirrorsays} to read off the exact type IIA  prepotential
$\mathcal{F} = \mathcal{F}_{\text{pert.}}+\mathcal{F}_{\text{inst.}}$
from the type IIB periods 
\eqref{eq:iibperdef}
expressed in a suitable integral symplectic basis
$\bigl\{[\Tilde{\alpha}^A],[\Tilde{\beta}_A] \bigr\}_{A=0}^{\Tilde{h}^{2,1}}$  of $H^3(\wt{X},\mathbb{Z})$.
To achieve this in practice, we will need to compute the period integrals in compactification on $\wt{X}$, and also compute the topological data of $X$ appearing in \eqref{eq:pert_prepotential}.

Both computations are feasible in the context of Calabi-Yau hypersurfaces in toric varieties.
In this section we recall key properties of such hypersurfaces, with particular attention to the gauge-invariant coordinates on complex structure moduli space.

\subsection{Polytopes and moduli}
Let $M\simeq \mathbb{Z}^4$ be a four-dimensional lattice, and let $N\simeq \mathbb{Z}^4$ be its dual lattice with respect to the Euclidean inner product $\langle \cdot,\cdot \rangle$. A \textit{lattice polytope} $\Delta$ is the convex hull of a finite set of lattice points in $M$.
A \textit{reflexive polytope} is a lattice polytope $\Delta\subset M$ whose polar dual 
\begin{equation}
\Delta^\circ\equiv \bigl\{p\in N|\, \langle p,q \rangle\geq -1\quad  \forall q\in \Delta \bigr\}
\end{equation}
is also a lattice polytope. Since $(\Delta^\circ)^\circ=\Delta$, the dual of a reflexive polytope is reflexive.

Let $\mathcal{T}$ be a fine, regular, and star triangulation of a reflexive polytope\footnote{For simplicity of presentation we will take $\Delta^{\circ}$ to be favorable.  Extending our results to non-favorable polytopes requires additional bookkeeping, but no new techniques.}
$\Delta^\circ$. Then, the cones over the simplices of $\mathcal{T}$ define a \textit{toric fan} $\Sigma_{\mathcal{T}}$ of a smooth toric fourfold $V_{\Delta^\circ,\mathcal{T}} \equiv V$.  
The one-dimensional cones (edges) in $\Sigma_{\mathcal{T}}$ are the cones over the lattice points $p\in \Delta^\circ$ other than the origin.  Given $n$ such points we will label them as $p^I$, $I=1,\ldots,n$. To each of the $p^I$ we associate a generator of the Cox ring of homogeneous coordinates $x_I$, and
these give rise to the toric divisors
\begin{equation}
    \hat{D}_I:=\{x_I=0\}\subset V_{\Delta^\circ,\mathcal{T}}\, ,
\end{equation}
and more generally a cone over a simplex in $\Sigma_{\mathcal{T}}$ defines a toric sub-variety of $V_{\Delta^\circ,\mathcal{T}}$ via the complete intersection of toric divisors associated with the edges of the cone. 

A $\mathbb{Z}$-basis of linear relations among the $p^I$ defines a GLSM matrix ${Q^a}_I$, $a=1,\ldots,n-4$. Each row of $Q$ encodes the scaling weight of the homogeneous coordinates $x_I$ under a $\mathbb{C}^*$ equivalence relation
\begin{equation}
    [x_1:\ldots:x_n]=\left[\prod_a\lambda_a^{{Q^a}_1}x_1:\ldots : \prod_a\lambda_a^{{Q^a}_n}x_n\right]\, ,\quad \forall \Vec{\lambda}\in (\mathbb{C}^*)^{n-4}\, .
\end{equation}
The toric divisor classes $[\hat{D}_I]$ generate $H^2(V_{\Delta^\circ,\mathcal{T}},\mathbb{Z})$, and can be written in terms of a basis $\{[\hat{H}_a\}_{a=1}^{h^{1,1}(V)}$ as
\begin{equation}
    [\hat{D}_I]=\sum_{a=1}^{n-4}{Q^a}_I [\hat{H}_a]\, ,
\end{equation}
and thus we have  $h^{1,1}(V)=n-4$.

We may (over-)parameterize the K\"ahler form on $V$ as
\begin{equation}
    J=\sum_I \mathfrak{t}^I [\hat{D}_I]\, ,
\end{equation}
using $h^{1,1}(V)+4$ redundant coordinates, or as 
\begin{equation}
    J=\sum_a t^a [\hat{H}_a]\, ,\quad t^a:={Q^a}_I \mathfrak{t}^I\, ,
\end{equation}
using $h^{1,1}(V)$ gauge-invariant coordinates on the K\"ahler cone of $V$.

The dual polytope $\Delta\equiv (\Delta^\circ)^\circ$ encodes the holomorphic monomials of the anti-canonical line bundle $\mathcal{O}_V(-K)$ as follows. We will denote the integer points in $\Delta$ other than the origin by $q^{\Tilde{I}}$, $\Tilde{I}=1,\ldots,m$.  Each of the $q^{\Tilde{I}}$ defines a monomial
\begin{equation}
    s_{\Tilde{I}}:=\prod_{I=1}^{n} x_I^{\langle p^I,q^{\Tilde{I}} \rangle+1}\, ,
\end{equation}
and we further define $s_0:=\prod_I x_I$, and
\begin{equation}
    S_{\Tilde{I}}:=\frac{s_{\Tilde{I}}}{s_0}\, .
\end{equation}
A GLSM charge matrix ${\Tilde{Q}^{\Tilde{a}}}_{~\Tilde{I}}$ for the dual polytope $\Delta$ defines multiplicative relations among the $s^{\Tilde{I}}$,
\begin{equation}\label{eq:multiplicative_relations_among_monomials}
    \prod_{\Tilde{I}} \left(S_{\Tilde{I}}\right)^{{\Tilde{Q}^{\Tilde{a}}}_{~\Tilde{I}}}=1\, ,\quad \forall \Tilde{a}=1,\ldots,m-4\, .
\end{equation}
A generic anti-canonical hypersurface $X$ in $V$ is the vanishing locus of a polynomial
\begin{equation}
    f=\Psi^0 S_0-\sum_{\Tilde{I}=1}^m \Psi^{\Tilde{I}}S_{\Tilde{I}}\, ,
\end{equation}
specified in terms of $m+1$ complex parameters $\Psi^{\Tilde{I}}$.
Such a hypersurface $X$ is a smooth Calabi-Yau \cite{Batyrev:1993oya}.

We thus see that points in $\Delta^\circ$ correspond to divisors of $V$, while points in the dual polytope $\Delta$ correspond to monomials of the anti-canonical line bundle. 
For our purposes, 
in both $\Delta$ and $\Delta^\circ$,
it suffices to consider points not interior to facets.
Toric divisors associated to points interior to facets of $\Delta^\circ$ do not intersect the generic Calabi-Yau hypersurface $X$, while part of the automorphism group of $V$ can be used to gauge fix $\Psi_{\Tilde{I}}=0$  for all $\Tilde{I}$ with $q^{\Tilde{I}}$ interior to a facet of $\Delta$. 

Even after this gauge fixing prescription, the coordinates $\Psi_{\Tilde{I}}$ still overparameterize the moduli space. 
First, the $\Psi_{\Tilde{I}}$ are at most projective coordinates on the space of inequivalent Calabi-Yau hypersurfaces, because polynomials differing only by an overall $\mathbb{C}^*$ scale lead to the same hypersurface. Second, non-trivial scaling symmetries acting on the toric coordinates lead to distinct choices of coefficients leading to  
hypersurfaces related by automorphisms.
For generic coefficients we can use 
\begin{equation}
    \tilde{\psi}^{\Tilde{a}}:=\prod_{\Tilde{I}} \left(\frac{\Psi^{\Tilde{I}}}{\Psi^0}\right)^{{\Tilde{Q}^{\Tilde{a}}}_{~\Tilde{I}}}\,,
\end{equation}
with $\tilde{a} = 1,\ldots ,h^{2,1}(X)$, 
as gauge-invariant local coordinates.  Globally, however, discrete symmetries of $V$ (or intrinsic symmetries of $X$) may still identify the threefolds $X$ arising from some distinct $\tilde{\psi}^{ \Tilde{a} }$.

Exchanging $\Delta^{\circ}\leftrightarrow \Delta$ in the above construction gives rise to the mirror dual $\wt{X}$, as shown by Batyrev. We denote by $\psi^a$ the gauge-invariant coordinates on the complex structure moduli space of $\wt{X}$, and the map between the moduli is
\begin{equation}
\frac{\log \psi^a}{2\pi i} =  u^a+\mathcal{O}(e^{2\pi i t^a})\, .
\end{equation}
For a detailed implementation of the mirror map, see \S\ref{sec:hktyandbeyond}.

\subsection{Example of a cubic in $\mathbb{P}^2$}\label{sec:cubic}

\begin{figure}
	\centering
	\begin{tabular}{c c}
		\includegraphics[keepaspectratio,width=5cm]{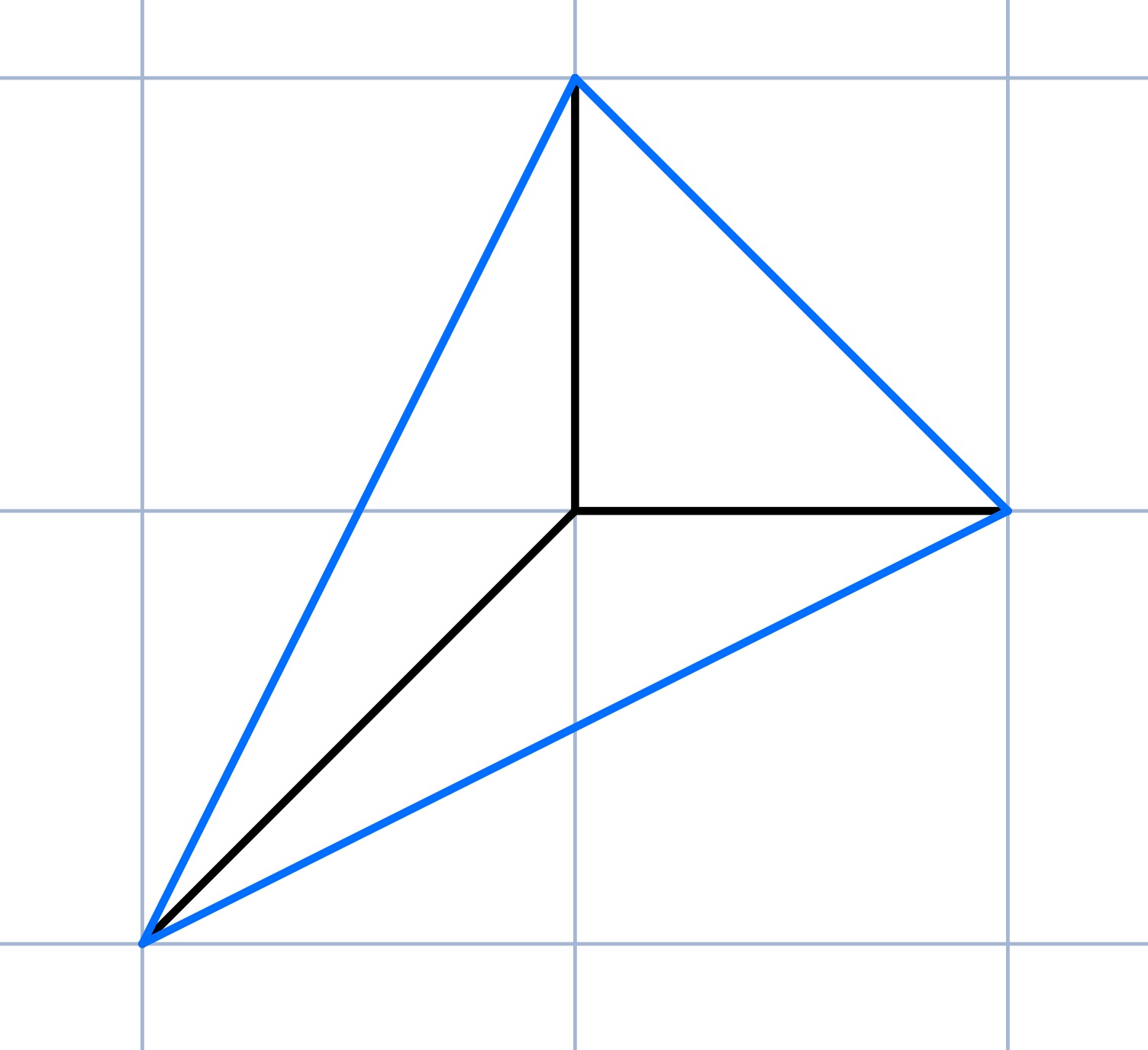} \hspace{1cm} &
		\includegraphics[keepaspectratio,width=5cm]{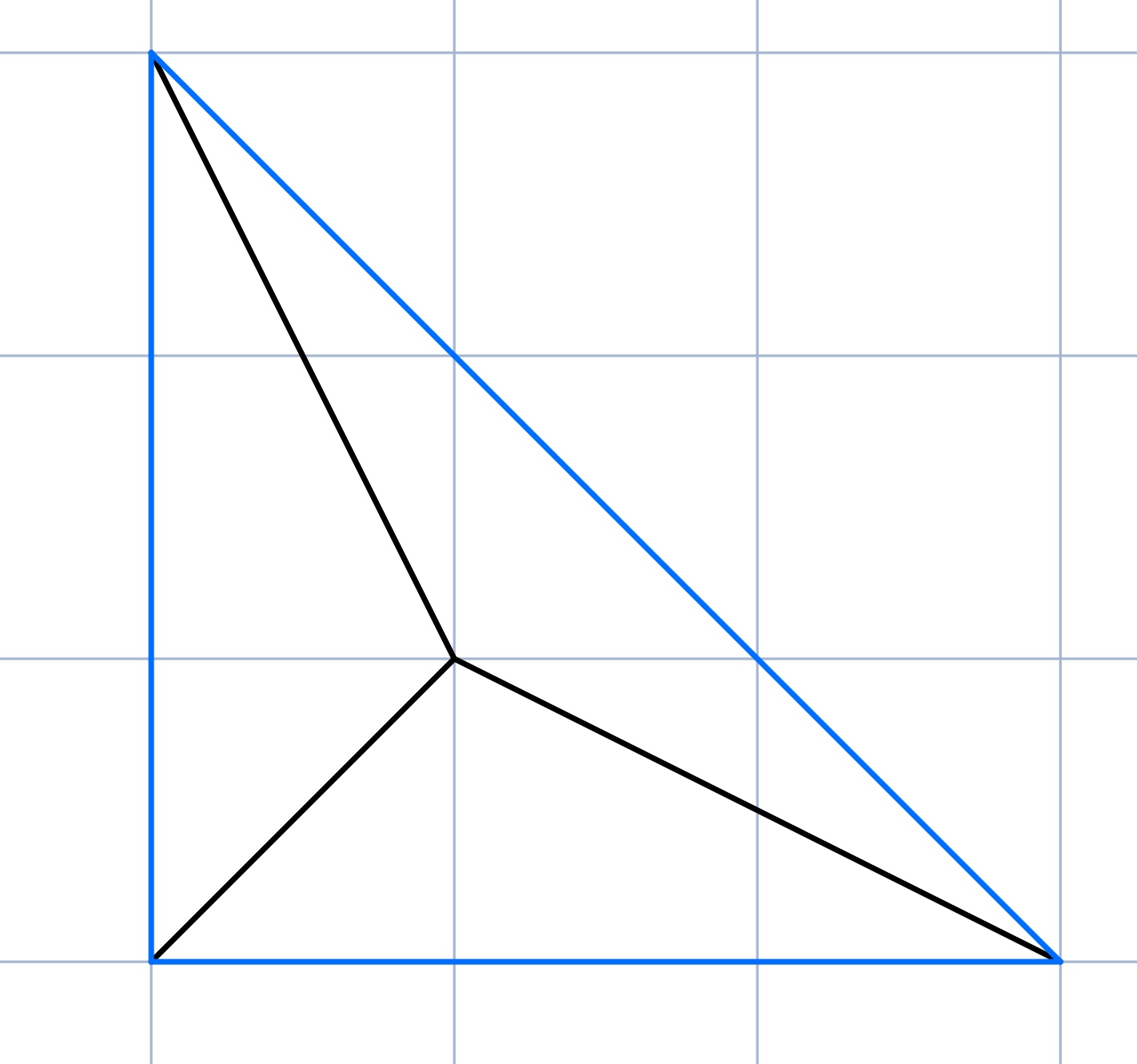}
	\end{tabular}
\caption{The fans (black) and convex hulls (blue) of $\mathbb{P}^2$ (left) and $\mathbb{P}^2/\mathbb{Z}_3$ (right).}
\label{fig:P2fan}
\end{figure}

Let us briefly exhibit the above with a very simple example: a cubic in $\mathbb{P}^2$. The edges of the toric fan of $\mathbb{P}^2$ are the cones over the points (see Fig.~\ref{fig:P2fan}) 
\begin{equation}
\begin{pmatrix}
p^1 & p^2 & p^3
\end{pmatrix}=\begin{pmatrix}
1 & 0 & -1 \\
0 & 1 & -1
\end{pmatrix}\, ,
\end{equation}
which satisfy $p^1+p^2+p^3=0$, so the GLSM charge matrix is $Q=(1,1,1)$. The convex hull over these points defines $\Delta^{\circ}$, and we have three homogeneous coordinates
\begin{equation}
    [x_1:x_2:x_3]=[\lambda x_1:\lambda x_2:\lambda x_3]\, ,\quad \forall \lambda\in\mathbb{C}^*\, .
\end{equation}
The dual polytope $\Delta\equiv (\Delta^{\circ})^\circ$ contains the points (see Figure \ref{fig:P2fan})
\begin{equation}
\begin{pmatrix}
q^1 & \cdots & q^9
\end{pmatrix}=\begin{pmatrix}
-1 & -1 &  2 & -1  & -1 & 0 & 1 & 0  & 1\\
 2  & -1 & -1 &  1  & 0  & 1 & 0 & -1 & -1
\end{pmatrix}\, ,
\end{equation}
as well as the origin. There are ten anti-canonical monomials
\begin{equation}
(s_0,\ldots,s_{9})=(x_0x_1x_2,x_2^{3},\ldots,x_1^2x_0)\longrightarrow (S_1,\ldots,S_{9})= \left(\frac{x_2^{2}}{x_0x_1},\ldots,\frac{x_1}{x_2}\right)\, .
\end{equation}
Starting with a general linear combination of these monomials, we may use up all but the diagonal part of the $PGL(3,\mathbb{C})$ symmetry group of $\mathbb{P}^2$ in order to set to zero the coefficients of all but the vertex monomials,
\begin{equation}\label{eq:cubic_in_P2}
    f\rightarrow \Psi_0 x_1x_2 x_3- \Psi_1 x_1^3-\Psi_2 x_2^3-\Psi_3 x_3^3\, .
\end{equation}
The vanishing locus defines an elliptic curve, and its complex structure is parameterized by a gauge invariant coordinate
\begin{equation}
    \psi:=\frac{\Psi_1\Psi_2\Psi_3}{\Psi_0^3}\, .
\end{equation}
Indeed, the complex structure $\tau$ of the elliptic curve defined by the vanishing of \eqref{eq:cubic_in_P2} is determined by the discriminant
\begin{equation}
    \text{Disc}(\psi)=1-27\psi\, ,
\end{equation}
via
\begin{equation}
    j(\tau)=\frac{27(8\,\text{Disc}-9)^3}{\text{Disc}^3(\text{Disc}-1)}\, ,
\end{equation}
where $j(\tau)$ is the modular invariant $j$-function.

The toric fan obtained by promoting the three vertices $(q^1,q^2,q^3)$ to edges of a toric fan is the fan of $\mathbb{P}^2/\mathbb{Z}_3$. As the remaining points on the boundary of $\Delta$ are interior to facets, the generic anti-canonical divisor is smooth, and is again an elliptic curve. Thus, as expected the mirror of an elliptic curve is again an elliptic curve.

\section{Computing Periods and the Prepotential}\label{sec:hktyandbeyond}

\subsection{Fundamental period}

On the type IIB side of the mirror map, the prepotential can be determined by computing the period integrals \eqref{eq:period2} in an integral symplectic basis.
For a Calabi-Yau hypersurface defined by the vanishing of a generic anti-canonical polynomial $f$ in a toric fourfold, it turns out to be straightforward to obtain the period integrals by computing a \textit{fundamental period}, as we will now explain.

We will consider a dual pair of reflexive polytopes $(\Delta,\Delta^\circ)$, and, following the conventions set in \S\ref{sec:batyrev}, we let $(\wt{\mathcal{T}},\mathcal{T})$ be FRSTs of these polytopes, and we denote by $(\wt{X},X)$ the respective generic Calabi-Yau hypersurfaces.

By definition the toric variety $\wt{V}$ contains a dense algebraic torus $(\mathbb{C}^*)^{4}$, parameterized by $\mathbb{C}^*$-valued coordinates $(\mathbbm{t}^1,\ldots,\mathbbm{t}^4)$. For a smooth simplicial toric fourfold $\wt{V}$ this is easy to make explicit: one considers the dense subset $\mathcal{U}:=\bigl\{[x_1:\ldots: x_n]\in \wt{V}|\, x_{\Tilde{I}}\neq 0\,\, \forall \Tilde{I}\bigr\}$, where one can gauge fix all the toric scaling relations. For example, one may pick a full-dimensional simplex $\sigma$ of the triangulation $\wt{\mathcal{T}}$, and set $x_q=1$ for all $q\notin \sigma$. The remaining four coordinates can be taken to be $(\mathbbm{t}^1,\ldots,\mathbbm{t}^4)$.

The holomorphic three-form $\wt{\Omega}$ of $\wt{X}$ can be written as the contour integral
\begin{equation}
    \wt{\Omega}=\oint_{f=0}\frac{d\mathbbm{t}^1\wedge d\mathbbm{t}^2\wedge d\mathbbm{t}^3\wedge d\mathbbm{t}^4}{(2\pi i)^4\cdot f(\mathbbm{t})}\, .
\end{equation}
Next, we fix an $\epsilon\ll 1$ and set $\mathbbm{t}^i=\epsilon\cdot e^{2\pi i \phi^i}$ for $i=1,2,3$, as a function of phases $\phi^i\in [0,1)$.  We restrict the coefficients of the polynomial $f$ to take values such that $f=0$ has a solution branch for $\mathbbm{t}^4(\mathbbm{t}^1,\mathbbm{t}^2,\mathbbm{t}^3)=\mathcal{O}(\epsilon)$. This solution branch defines a $T^3\subset \wt{X}$, which  can be thought of as the SYZ fiber \cite{Strominger:1996it}.
The integral of $\wt{\Omega}$ over this three-torus can be written as a contour integral over the vanishing loci of the $\mathbbm{t}^i$:
\begin{equation}\label{eq:fundef}
\varpi_0(\psi):=\Psi^0\int_{T^3}\wt{\Omega}=\Psi^0\oint_{|\mathbbm{t}^1|=\epsilon}\frac{d\mathbbm{t}^1}{2\pi i}\cdots \oint_{|\mathbbm{t}^4|=\epsilon} \frac{d\mathbbm{t}^4}{2\pi i}\frac{1}{f(\mathbbm{t})}\, .
\end{equation}
The relation \eqref{eq:fundef} defines the \emph{fundamental period} $\varpi_0(\psi)$.

Using the residue theorem one can evaluate this expression. Namely, we have
\begin{equation}
\varpi_0(\psi)=\oint\frac{d\mathbbm{t}^1}{2\pi i \mathbbm{t}^1}\cdots \oint \frac{d\mathbbm{t}^4}{2\pi i \mathbbm{t}^4}\frac{1}{1-\sum_i \tfrac{\Psi^I}{\Psi^0} S_I} = \sum_{k=0}^\infty \left.\left(\sum_I \tfrac{\Psi^I}{\Psi^0} S_I\right)^k\right|_{\mathbbm{t}\text{-independent}}\, ,
\end{equation}
where on the right-hand side for each $k$ one keeps only the constant terms in the expansion of $\left(\sum_I \tfrac{\Psi^I}{\Psi^0} S_I\right)^k$.

Applying the multinomial theorem one further simplifies the fundamental period to
\begin{align}
\varpi_0(\psi)= \sum_{k_1=0}^\infty\cdots \sum_{k_{h^{2,1}+4}=0}^\infty \frac{(\sum_I k_I)!}{\prod_I k_I!}\begin{cases}
    \prod_{I=1}^{h^{2,1}+4}\left(\tfrac{\Psi^I}{\Psi^0}\right)^{k_I} & \text{if}\quad k_I q^I=0\\
    0 & \text{otherwise}
\end{cases}\, .\end{align}
where we used that by virtue of \eqref{eq:multiplicative_relations_among_monomials}, $\prod_I S_I^{k_I}=1$ if and only if the vector $k_I$ defines a linear relation $\sum_I k_Iq^I=0$ among the points $q^I$ of the dual polytope $\Delta^\circ$.

But a linear relation among the points in $\Delta^\circ$ defines a curve class in the mirror toric variety $V$ (defined by an FRST of $\Delta^{\circ}$), and in its corresponding Calabi-Yau hypersurface $X$, so instead we may sum over curve classes in $H^2(V,\mathbb{Z})$ defined by the GLSM matrix
\begin{equation}
    k_I=\sum_{a=1}^{h^{1,1}(V)}n_a{Q^a}_I\, ,
\end{equation}
subject to the constraint that the $k_I\geq 0$. In other words one sums over all curve classes in $V$ that have non-negative intersection with all the toric divisors $D_I$. This identifies the relevant curve classes as elements of the \emph{cone of movable curves} $\text{Mov}_V$ in the toric ambient variety $V$, and in particular all such curves are effective in $V$.

Thus, we may express the fundamental period as
\begin{equation}\label{eq:fundamental_period_final}
\varpi_0(\psi)=\sum_{\vec{n}\in \text{Mov}_V\cap H^2(V,\mathbb{Z})}\frac{\Gamma(1+n_a {Q^a}_0)}{\prod_I \Gamma(1+n_a {Q^a}_I)}\psi^{\vec{n}}=: \sum_{\vec{n}}c_{\vec{n}}\psi^{\vec{n}}\, ,
\end{equation}
where ${Q^a}_0:=\sum_I {Q^a}_I$ is the anti-canonical divisor, we have replaced $k!\rightarrow \Gamma(1+k)$ for later convenience,
and we have defined
\begin{equation}
\psi^{\vec{n}} := \prod_{a=1}^{h^{1,1}(V)} (\psi^a)^{n_a}\,.
\end{equation}

After writing the fundamental period as in \eqref{eq:fundamental_period_final} one can extend the sum to run over all effective curve classes in $H^2(V,\mathbb{Z})$, i.e.~over classes
$\vec{n}\in \mathcal{M}_V\cap H^2(V,\mathbb{Z})$, with $\mathcal{M}_V$ the Mori cone\footnote{To compute the Mori cone of a toric variety $V$ we use the 
Oda-Park algorithm \cite{oda1991linear,Berglund:1995gd}.} of $V$.
Let us explain why this is possible.
Any effective curve $\vec{n}\in \mathcal{M}_V$
has a non-negative intersection with the anti-canonical divisor, and so the numerator of $c(\vec{n})$ remains regular.  However, for $\vec{n}\in \mathcal{M}_V \setminus \text{Mov}_V$, i.e.~for curve classes in the complement of $\text{Mov}_V$ within the cone of effective curves, the denominator develops a pole.
Thus the sum
in 
\eqref{eq:fundamental_period_final}
remains unchanged upon extending the range from 
$\text{Mov}_V$  to $\mathcal{M}_V$.

\subsection{Picard-Fuchs system}
We now turn to the Picard-Fuchs system, which is a system of linear partial differential equations that is equivalent to the first order Gauss-Manin system of a Calabi-Yau manifold \cite{griffiths1968periods1,griffiths1968periods2}. 

Due to the automorphisms of $\wt{V}$, not just the fundamental period $\varpi_0$, but in fact all the period integrals,  
can be written as
\begin{equation}\label{eq:normalized_period_ansatz}
    \Vec{\Pi}(\Psi^0,\Psi^I)=:\frac{1}{\Psi^0} \Vec{\varpi}(\psi)\, ,
\end{equation}
in terms of some functions $\vec{\varpi}$ that depend only on the gauge-invariant coordinates $\psi^a$.

The multiplicative relations among the monomials \eqref{eq:multiplicative_relations_among_monomials} imply that the period vector satisfies a set of differential equations called the generalized Gelfand-Kapranov-Zelevinksy (GKZ) hypergeometric system \cite{GKZ:1989}. Concretely, for a choice of basis of linear relations such that ${Q^a}_I\geq 0$, one finds
\begin{equation}\label{eq:pf1}
    \left[\left(-\frac{\del}{\del \Psi^0}\right)^{{Q^a}_0}-\prod_{I} \left(\frac{\del}{\del \Psi^I}\right)^{{Q^a}_I}\right] \Vec{\Pi}(\Psi^0,\Psi^I)=0\,.
\end{equation}
Using \eqref{eq:normalized_period_ansatz} one can rewrite \eqref{eq:pf1} as the condition that a set of $\Tilde{h}^{2,1}$ differential operators $\mathcal{L}_a$ in the $\psi^a$ annihilate the normalized periods $\vec{\varpi}$,
\begin{equation}\label{eq:gen_GKZ}
    \hat{\mathcal{L}}_a\Vec{\varpi}(\psi)=0\, ,\quad \forall a=1,\ldots,\Tilde{h}^{2,1}\, ,
\end{equation} 
where
\begin{equation}
    \hat{\mathcal{L}}_a:=\psi^a  (\hat{\Theta}_0+{Q^a}_0)\times \ldots \times (\hat{\Theta}_0+1)-\prod_{I} {X^a}_I\, ,
\end{equation}
in terms of
\begin{equation}
    {X^a}_I:=(\hat{\Theta}_I-({Q^a}_I-1))\times \ldots \times (\hat{\Theta}_I-1)\times \hat{\Theta}_I\, ,
\end{equation}
with $\hat{\Theta}_I:=\sum_a {Q^a}_I \theta_a$, $\hat{\Theta}_0:=\sum_a {Q^a}_0 \theta_a$, and with logarithmic derivatives 
\begin{equation}
\theta_a:=\tfrac{\del}{\del \log(\psi^a)}\,.
\end{equation}

While all the periods satisfy this system of differential equations, the general solution of \eqref{eq:gen_GKZ} cannot be written as a linear combination of periods, i.e.~the rank of \eqref{eq:gen_GKZ} is larger than $2\Tilde{h}^{2,1}+2$.
However, one could try to  factor the differential operators to find a lower rank system of differential operators $\mathcal{L}_a$ such that the general solution of
\begin{equation}\label{eq:factorized_GKZ}
    \mathcal{L}_a\Vec{\varpi}(\psi)=0\, ,\quad \forall a=1,\ldots,\Tilde{h}^{2,1}\, ,
\end{equation}
is a linear combination of periods \cite{HKTY,HKTY2}. 
This direct approach has two limitations.
The first is that computing the Picard-Fuchs system by reducing the generalized GKZ system is computationally challenging and quickly becomes intractable even for modest numbers of moduli.  The second is that in some cases the GKZ system must be supplemented with additional data that can be difficult to compute \cite{HKTY}.  

One would therefore like to bypass the computation of the Picard-Fuchs system in determining period integrals. An elegant solution to this problem provided by \cite{HKTY,HKTY2} is as follows. One considers a special point $u^*$ in moduli space around which the monodromy is maximally unipotent \cite{morrison1993compactifications}: this is called a large complex structure (LCS) point. At $u^*,$ all but one of the linear combinations of the period integrals degenerate. Said differently, at $u^*$ the Picard-Fuchs system reduces to the principal part of the Picard-Fuchs system. As proven in \cite{HKTY2}, because the solutions to the Picard-Fuchs system maximally degenerate at $u^*,$ a basis of solutions 
expanded around 
$u^*$ can be obtained via the Frobenius method. This result relies on the structure of the Picard-Fuchs system and the existence of points of maximal unipotent monodromy. 

Unfortunately, a direct computation of the principal part of the Picard-Fuchs system is   not substantially easier than computing the entire Picard-Fuchs system.
The key to a direct computation of the periods is mirror symmetry, which equates LCS points with large volume limits of mirror dual Calabi-Yau manifolds \cite{Candelas:1990rm,Strominger:1996it}. In particular, the LCS monodromies are identical to the large volume monodromies that we have laid out in detail in \S\ref{sec:iiapert}, which in turn are determined entirely by classical geometric data --- in particular the intersection form --- of the mirror threefold.
While computing intersection forms of Calabi-Yau threefolds with many moduli has long been inaccessible, this problem has recently been overcome \cite{Demirtas:2018akl,Demirtas:2022hqf}, making the computation of the required classical geometric data readily available.  

With these tools in hand, we can write the general Calabi-Yau periods $\vec{\varpi}$ around an LCS point  
as suitable linear combinations of the fundamental period $\varpi_0$, and of the logarithmic functions
\begin{align} \label{eq:pfrelations}
\varpi^a(\psi)=\sum_{\vec{n}\in \mathcal{M}_V}\left. \frac{\del_{\rho_a}}{2\pi i}\left(c_{\vec{n}+\Vec{\rho}}\,\psi^{\vec{n}+\Vec{\rho}}\right)\right|_{\Vec{\rho}=0}&\, ,\quad \varpi^{ab}(\psi)=\sum_{\vec{n}\in \mathcal{M}_V}\left.\frac{\del_{\rho_a}\del_{\rho_b}}{(2\pi i)^2}\left(c_{\vec{n}+\Vec{\rho}}\,\psi^{\vec{n}+\rho}\right)\right|_{\Vec{\rho}=0}\,,\nonumber\\
\varpi^{abc}(\psi)=\sum_{\vec{n}\in \mathcal{M}_V}&\left. \frac{\del_{\rho_a}\del_{\rho_b}\del_{\rho_c}}{(2\pi i)^3}\left(c_{\vec{n}+\Vec{\rho}}\,\psi^{\vec{n}+\Vec{\rho}}\right)\right|_{\Vec{\rho}=0}\, .
\end{align}
Expanding these around the LCS point at zeroth order in $\psi$ one obtains
\begin{align}
&\varpi_0\simeq 1\, ,\quad  \varpi^a\simeq \frac{\log(\psi^a)}{2\pi i}\, ,\quad
\frac{1}{2}\kappa_{abc}\varpi^{ab}\simeq \frac{1}{2}\kappa_{abc}\varpi^a \varpi^b-\frac{1}{24}c_a\nonumber\\
&\frac{1}{3!}\kappa_{abc}\varpi^{abc}\simeq \frac{1}{3!}\kappa_{abc}\varpi^a \varpi^b \varpi^c-\frac{1}{24}c_a \varpi^a+\frac{\zeta(3)}{(2\pi i)^3}\chi\, ,
\end{align}
which can be verified using the fact that
\begin{align}
c_a=&=\frac{1}{2}\kappa_{abc}\Bigl({Q^b}_0{Q^c}_0-\sum_I {Q^b}_I {Q^c}_I\Bigr)\, ,\\
\chi=&-\frac{1}{3}\kappa_{abc}\Bigl({Q^a}_0 {Q^b}_0 {Q^c}_0-\sum_I {Q^a}_I {Q^b}_I {Q^c}_I\Bigr)\, ,
\end{align}
as follows from the adjunction formula.
Comparing with the period vector obtained from the prepotential \eqref{eq:pert_prepotential}, one concludes that
\begin{equation}\label{eq:period2}
\Pi(\psi)=\frac{1}{\varpi_0}\begin{pmatrix}
\frac{1}{3!}\kappa_{abc}\varpi^{abc}+\frac{1}{12}c_a \varpi^a\\
-\frac{1}{2}\kappa_{abc}\varpi^{bc}+a_{ab}\varpi^b\\
\varpi_0\\
\varpi^a
\end{pmatrix}\, ,
\end{equation}
where we have used the freedom to rescale $\wt{\Omega}(\psi)$ to fix the overall normalization of $\Pi(\psi)$ so that it matches the form one gets from a prepotential, in the conventions used in 
\eqref{eq:iiaper}.
Importantly, the boundary data at LCS is sufficient to fix all integration constants, as shown in \S\ref{sec:iiapert}.

\subsection{Periods for a cubic in $\mathbb{P}^2$}

Let us illustrate the method of \cite{HKTY} by continuing to treat the example of a cubic in $\mathbb{P}^2$, as in \S\ref{sec:cubic}.
First, the fundamental period reads
\begin{equation}
\varpi_0(\psi)=\sum_{n=0}^{\infty} \frac{\Gamma(1+3n)}{ \Gamma(1+n)^3}\psi^n=\left._2 F_1\right.\left(\frac{1}{3},\frac{2}{3},1,1-\mathrm{Disc}\right)\, ,
\end{equation}
where $\mathrm{Disc}$ denotes the discriminant, 
and the GKZ differential operator is
\begin{equation}
    \hat{\mathcal{L}}=3\psi (3\theta+3)(3\theta+2)(3\theta+1)-\theta^3=\theta\times \left(3\psi(3\theta+2)(3\theta+1)-\theta^2\right)\equiv \theta\times \mathcal{L}\, .
\end{equation}
The operator $\mathcal{L}$ defined via the above factorization annihilates the general period, and defines the hypergeometric differential equation of degree $(\frac{1}{2},\frac{2}{3},1)$. The logarithmic solution defined by applying the prescription \eqref{eq:pfrelations} is indeed the second solution to $\mathcal{L}\Vec{\varpi}=0$, 
\begin{align}
    \varpi^1(\psi)=\sum_{n=0}^{\infty}\left. \frac{\del_{\rho}}{2\pi i}\left(c(n+\rho)\psi^{n+\rho}\right)\right|_{\rho=0}&=\frac{\log(\psi)}{2\pi i}\varpi_0(\psi)+3\sum_{n=0}^{\infty}\frac{(3n)!\cdot (H_{3n}-H_n)}{(n!)^3}\psi^n\nonumber\\
    &=\frac{i}{\sqrt{3}}\left._2 F_1\right.\left(\frac{1}{3},\frac{2}{3},1,\mathrm{Disc}\right)\, ,
\end{align}
where $H_n$ is the $n$th harmonic number.

Normalizing the period vector of the elliptic curve as
\begin{equation}
    \Vec{\Pi}=\begin{pmatrix}
        \tau\\
        1
    \end{pmatrix}:=\begin{pmatrix}
         \varpi^1(\psi)/ \varpi_0(\psi)\\
        1
    \end{pmatrix}\, ,
\end{equation}
one then recovers a well-known expression for the inverse of the $j$-function,
\begin{equation}
    \tau=\frac{i}{\sqrt{3}}\frac{\left._2 F_1\right.\left(\frac{1}{3},\frac{2}{3},1,\mathrm{Disc}\right)}{\left._2 F_1\right.\left(\frac{1}{3},\frac{2}{3},1,1-\mathrm{Disc}\right)}\, .
\end{equation}

\subsection{Enumerative invariants}\label{subsec:enumerative_invariants}
 
The first conclusion we draw from \eqref{eq:period2} is that the flat coordinates $u^a$ can be expanded for suitable $\psi^a$ as
\begin{equation}
    u^a=\frac{\varpi^a}{\varpi_0}=\frac{\log(\psi^a)}{2\pi i}+\frac{1}{2\pi i }  \frac{c^a(\psi)}{\varpi_0}\, ,
\end{equation}
where we define power series $c^a(\psi)$ via the perturbative expansion 
\begin{equation}\label{eq:cadef}
    c^a(\psi):=\sum_{\Vec{n}\in \mathcal{M}_V}  c^a_{\Vec{n}}\,\psi^{\Vec{n}}\, ,\quad \text{with} \quad c^a_{\Vec{n}}:=\del_{\rho_a}c_{\Vec{n}+\Vec{\rho}}\Bigl|_{\vec{\rho}=0}\, .
\end{equation}

Similarly, we define
\begin{equation}\label{eq:cabdef}
    c^{ab}(\psi):=\sum_{\Vec{n}\in \mathcal{M}_V} c^{ab}_{\Vec{n}} \,\psi^{\Vec{n}}\, ,\quad \text{with} \quad c^{ab}_{\Vec{n}}:=\del_{\rho_a}\del_{\rho_b}c_{\Vec{n}+\Vec{\rho}}\Bigr|_{\vec{\rho}=0}\, .
\end{equation}
In terms of these variables we may write the periods as
\begin{equation}\label{eq:periodf}
    \mathcal{F}_a=-\frac{1}{2}\kappa_{abc}\frac{\varpi^{bc}}{\varpi_0}+a_{ab}\frac{\varpi^b}{\varpi_0}=-\frac{1}{2}\kappa_{abc}u^bu^c+a_{ab}u^b+\frac{c_a}{24}-\frac{1}{2(2\pi i)^2}\kappa_{abc}\frac{\hat{c}^{bc}-c^bc^c}{\varpi_0}\, ,
\end{equation}
where   
\begin{align}\label{eq:chat}
    \hat{c}^{ab}(\psi)&:=\sum_{\Vec{n}\in \mathcal{M}_V}\hat{c}^{ab}_{\Vec{n}}\psi^{\vec{n}}:=\sum_{\Vec{n}\in \mathcal{M}_V}\left(c^{ab}_{\Vec{n}}-\frac{\pi^2}{6}\Bigl[{Q^a}_0{Q^b}_0-\sum_I {Q^a}_I{Q^b}_I\Bigr]c_{\Vec{n}}\right)\psi^{\vec{n}}\nonumber\\
    &\equiv c^{ab}(\psi)-\frac{\pi^2}{6}\Bigl[{Q^a}_0{Q^b}_0-\sum_I {Q^a}_I{Q^b}_I\Bigr]\varpi_0\, .
\end{align}
Comparing with \eqref{eq:FinstIIA}, we find 
\begin{equation}
\label{eq:perturbative_GV_formula}
    \boxed{~\sum_{\vec{n}\in \mathcal{M}_X}n_a\text{GV}^0_{\vec{n}}\,\text{Li}_2\biggl(\psi^{\Vec{n}}\,\mathrm{exp} \Bigl({ n_b \tfrac{c^b(\psi)}{\varpi_0(\psi)}}\Bigr)\biggr)=\phantom{\Biggl(}\frac{1}{2}\kappa_{abc}\frac{\hat{c}^{bc}(\psi)-c^b(\psi)c^c(\psi)}{\varpi_0(\psi)}\,,~}
\end{equation}
where we have restricted the sum to run over
$\vec{n} \in \mathcal{M}_X \subset \mathcal{M}_V$, as $\text{GV}^0_{\vec{n}}=0$ for $\vec{n}$ outside $\mathcal{M}_X$.

The master formula \eqref{eq:perturbative_GV_formula} can be evaluated on both sides order by order in the $\psi^a$, thus determining all the GV invariants systematically. Notably, the series coefficients of the relevant power series for $\varpi(\psi),c^a(\psi)$ and $\hat{c}^{ab}(\psi)$ are all rational.\footnote{Formulas for the $c_{\Vec{n}}$ and $\hat{c}^{ab}_{\Vec{n}}$ are given in Appendix \ref{sec:Appendix_cformulas}, following, with some minor simplifications, Appendix A of \cite{HKTY}.}

A few comments are in order: we have written all period components as sums over curve classes in the Mori cone $\mathcal{M}_V$ of the ambient toric variety $V$, but the formulas for $\varpi_0$ and $\varpi^a$ make no reference to a choice of triangulation $\mathcal{T}$ of the polytope $\Delta^\circ$,  which determines the toric fan of $V$ and thereby $\mathcal{M}_V$. Therefore, one expects that only curve classes in the intersection of \emph{all} the Mori cones --- associated with the set of toric varieties obtained via \emph{all} possible FRSTs of $\Delta^\circ$ --- can contribute. Indeed, curves in $\mathcal{M}_V$ that shrink across bistellar flips in $V$ 
have negative intersection with two of the toric divisors, leading to a double pole in the denominator of $c(\Vec{n})$. Therefore, these curves can at most contribute to $\varpi^{ab}$ and $\varpi^{abc}$, but not to $\varpi^a$ and $\varpi_0$. 
The periods $\mathcal{F}_a$ in \eqref{eq:periodf} do depend on the choice of triangulation of $\Delta^\circ$, both through the intersection form $\kappa_{abc}$, as well as through the set of curves that contribute to $\varpi^{ab}$.
 
Next we turn to a technical but computationally important point regarding the sum over curves in $\mathcal{M}_V$: on the one hand it turns out that the Mori cone $\mathcal{M}_V$ of an ambient variety $V$ is often much larger than the Mori cone of its Calabi-Yau hypersurface $\mathcal{M}_X$, so evaluating \eqref{eq:perturbative_GV_formula} over many points in $\mathcal{M}_V$  can come a significant computational cost while only predicting very few non-vanishing GV invariants. On the other hand, many birational morphisms between toric varieties associated with different triangulations of the polytope $\Delta^\circ$ 
do not actually affect the generic Calabi-Yau hypersurface.

We can make use of this fact and evaluate polynomials only within the intersection of $\mathcal{M}_V$ with all the Mori cones $\mathcal{M}_{V'}$ of toric varieties $V'$ that are birational to $V$ and such that the generic Calabi-Yau hypersurface remains smooth across the birational transformation $V\rightarrow V'$. We will denote this cone by $\mathcal{M}_V^{\cap}$. 
Typically one finds that $\mathcal{M}_V^{\cap}$ is a decent outer approximation to $\mathcal{M}_X$, and in all cases $\mathcal{M}_X \subseteq \mathcal{M}_V^{\cap}$.
Computing $\mathcal{M}_V^{\cap}$ directly in geometries with exponentially many phases is infeasible, but one of the authors has devised an effective algorithm for obtaining  $\mathcal{M}_V^{\cap}$ \cite{AndresThesis} that underlies the overall success of our method at large numbers of moduli.

\subsection{Consistent truncation of instanton spectrum}\label{sec:truncation}

The strategy will be to evaluate equation \eqref{eq:perturbative_GV_formula} order by order in expansion in the $\psi^a$.
In order to do this efficiently and consistently, one has to truncate the semi-group of effective curves in $\mathcal{M}^{\cap}_V$ to a finite set of curves, which truncates the infinite power series in the $\psi^a$ to polynomials. 

In order to compute the GV invariants of a set of curves $\mathcal{S}$, one has to be a bit careful about the truncation scheme. Evaluating both sides of \eqref{eq:perturbative_GV_formula} can essentially be split into two steps: 
\begin{enumerate}
    \item Computing the coefficients $c_{\vec{n}}$,  
$c^a_{\Vec{n}}$, and 
$c^{ab}_{\Vec{n}}$ of monomials in $\varpi_0(\psi)$, $c^a(\psi)$ and $c^{ab}(\psi)$, cf.~\eqref{eq:fundamental_period_final},\eqref{eq:cadef}, \eqref{eq:cabdef}.

    \item Evaluating basic functions thereof.
\end{enumerate}
The resulting coefficient of a monomial associated with a curve class $\Vec{n}\in \mathcal{M}_V^{\cap}$ in the expansion of \eqref{eq:perturbative_GV_formula} generically depends on all the coefficients of the output of step (a) ---  i.e. the coefficients $c_{\Vec{n}'}$, $c^a_{\vec{n}'}$ and $c^{ab}_{\Vec{n}'}$ --- for all curve classes $\Vec{n}'\in \mathcal{M}_V$ such that
\begin{equation}
    \Vec{n}=\Vec{n}'+\Vec{n}''\, ,
\end{equation}
for some $\Vec{n}''\in \mathcal{M}_V^{\cap}$. Therefore, truncating all power series in $\psi^a$ to a finite set $\mathcal{S}$ of curves in $\mathcal{M}_V^{\cap}$ leads to a consistent perturbative scheme determining  the GV invariants of all curves in $\mathcal{S}$ if and only if $\mathcal{S}$ is subject to a sort of `causality' constraint: for a point $\vec{n}\in \mathcal{M}_V^{\cap}$ we can define its \emph{causal diamond} 
\begin{equation}
    \lozenge_V(\Vec{n}):=\mathcal{M}_{V}^\cap \bigcap \left(\Vec{n}-\mathcal{M}_{V}^\cap\right)\, ,
\end{equation}
where $\Vec{n}-\mathcal{M}_{V}^\cap$ is minus the cone $\mathcal{M}_{V}^\cap$ translated by $\Vec{n}$. The causality constraint is that
\begin{equation}\label{eq:causality_in_Mori}
    \lozenge_V(\Vec{n})\subset \mathcal{S}\, ,\quad \forall \Vec{n}\in \mathcal{S}\, .
\end{equation}
One simple way to pick such a region $\mathcal{S}$ is by letting $\mathcal{S}=\lozenge_V(\Vec{n}_0)$ for some fixed choice of $\Vec{n}_0\in \mathcal{M}_V^{\cap}$. We will call this method the \emph{past light cone} method.

Another simple possibility is to choose an integer \emph{grading vector} $\vec{d}$ in the strict interior of the K\"ahler cone, i.e.
\begin{equation}
    \Vec{d}\in \Bigl(\mathcal{K}_V^{\cup}-\del \mathcal{K}_V^{\cup}\Bigr)\cap H^2(V,\mathbb{Z})\, ,
\end{equation}
where $\mathcal{K}_V^{\cup}$ is the cone dual to $\mathcal{M}_V^{\cap}$. Such a grading vector defines a semi-group homomorphism that we will call the \emph{degree} of a curve,
\begin{equation}
    \text{deg}_{\Vec{d}}:\,\mathcal{M}_{V}^\cap\rightarrow \mathbb{N}\, ,\quad \Vec{n}\mapsto \langle\Vec{n},\Vec{d}\rangle\, .
\end{equation}
We then choose an integer cutoff degree $\ell\in \mathbb{N}$.
The semi-group homomorphism property guarantees that for each curve $\Vec{n}$ with degree $d_{\Vec{n}}\leq \ell$, all sets of effective curves that can be added to yield $\Vec{n}$ have degrees $\leq d_{\Vec{n}}\leq \ell$ and are thus automatically included in the computation scheme. In other words, truncating all power series to effective curves with degree below any given cutoff $\text{deg}_{\Vec{d}}(\Vec{n})\leq \ell$ defines a set of curve classes that satisfies the causality constraint \eqref{eq:causality_in_Mori}, and one may set
\begin{equation}
    \mathcal{S} \rightarrow \mathcal{S}_{\Vec{d},\ell}:=\big\{\vec{n}\in \mathcal{M}_V^{\cap}|\, \text{deg}_{\Vec{d}}(\Vec{n})\leq \ell\bigr\}\, .
\end{equation}
We will refer to this method as the \emph{degree method}.

The degree method is particularly useful for approximating the prepotential, evaluated at some point $\vec{u}_0$ in moduli space, to fixed precision $\epsilon$: one first finds a nearby rational point $\Vec{d}'\approx \text{Im}(\Vec{u}_0)$, and clears denominators $\Vec{d}=k\cdot \Vec{d}'$ with $k \in \mathbb{N}$.  One then uses $\Vec{d}$ as the grading vector, taking a cutoff degree
\begin{equation}
    \ell =  \left\lceil k\cdot \frac{\log(1/\epsilon)}{2\pi} \right\rceil\, .
\end{equation}

The most general consistent truncation scheme is generated by the causal diamonds of any finite set of curves $\mathcal{S}'$, i.e.
\begin{equation}
    \mathcal{S}=\bigcup_{\Vec{n}\in \mathcal{S}'}\lozenge_V(\Vec{n})\, .
\end{equation}
 
The number of points in $\mathcal{M}_V^{\cap}$ up to some cutoff degree $\ell$ scales as the volume of the region,  
\begin{equation}
    \#(\mathcal{S}_{\Vec{d},\ell}) \sim \ell^{h^{1,1}}\, .
\end{equation}
Consider for instance a smooth simplicial cone $\mathcal{M}_{V}^\cap$ and a curve $\Vec{n}_0=(k,\ldots,k)$ for some $k>0$. Then, 
\begin{equation}
    \#\Bigl(\lozenge_V(\Vec{n}_0)\Bigr)=(1+k)^{h^{1,1}}\, ,
\end{equation}
which, for the largest Hodge number $h^{1,1}=491$, is already of order $6\times 10^{147}$ for $k=1$. 
Thus, for large $h^{1,1}$, computing GV invariants to parametrically large cutoff degree is exponentially expensive.
Nevertheless, we will be able to compute large numbers of nonzero GV invariants, even at $h^{1,1}=491$, after making suitable choices of grading vectors and working to modest cutoff degrees.
 
Alternatively, one can choose $\Vec{n}_0$ to lie in a low-dimensional face $m$ of $\mathcal{M}_V^\cap$. Then the causal diamond $\lozenge_V(\Vec{n})$ lies entirely in $m$ and the computational cost of computing GV invariants within $m$ scales as
\begin{equation}
    \#\Bigl(\lozenge_V(\ell\cdot \Vec{n}_0)\Bigr)\sim \ell^{\text{dim}(m)}\, .
\end{equation}
We would like to highlight the special case where $\vec{n}_0$ is proportional to an extremal ray of the Mori cone, i.e.~$\text{dim}(m)=1$. In this case, even at very large $h^{1,1}$, $\lozenge_V(\Vec{n}_0)$ is a sequence of curves on the extremal ray, and the computation of GV invariants along it is roughly as expensive as the computation of GV invariants at $h^{1,1}=1$.
 
In \S\ref{sec:examples} we give examples that illustrate each of the above methods. 

\section{Computational Algorithm}\label{sec:algorithm}

Thus far we have described an extension of the ideas of HKTY \cite{HKTY} that in principle allows one to compute the prepotential in an arbitrary mirror pair of Calabi-Yau hypersurfaces.
We now turn to a practical implementation.

The best-developed public software for performing the mirror map along the lines of
\cite{HKTY} 
is the {\tt{Instanton}} package for {\texttt{Mathematica}}, developed by Klemm and   Kreuzer \cite{instantonweb}. 

Apart from the limitations inherent to the original HKTY procedure, {\tt{Instanton}} also struggles to compute high-degree GV invariants, to handle Calabi-Yaus with a large number of moduli, and to make use of parallelization capabilities present in modern systems. For these reasons we rewrote the entire procedure in the \texttt{C++} programming language, making a number of functional improvements.  

Let us describe how the algorithm works.
Given a Calabi-Yau hypersurface, a grading vector $\vec{d}$, and a choice of maximum degree $\ell$, we start by constructing the full list of points (i.e.~monomials) whose degree is no larger than $\ell$. The monomials are sorted by degree and arranged in a list. We also construct a hash map\footnote{A \emph{hash map}, also known as a hash table or dictionary, is a data structure that stores pairs of \emph{keys} and \emph{values}, and performs the mapping from keys to values using a so-called \emph{hash function}. This function converts keys into indices that directly indicate the location of the corresponding value, which results in very efficient evaluation by avoiding iterations over the set of values.} that maps each monomial to the corresponding index of the list. Let us denote this hash map $\mathfrak{M}$ for future reference.

A central part of the algorithm is the efficient handling of polynomials with a given truncation. Whereas additions and subtractions are simple operations, polynomial multiplication is an expensive operation of time complexity $\mathcal{O}(n^2)$. Empirically, one finds that most polynomials are appreciably sparse, making multiplication using Fast Fourier Transforms (FFTs) impractical. Further complications include the high dimensionality and the extraordinarily large coefficients that appear. These large numbers need to be stored either as multi-precision floating-point numbers with \texttt{MPFR} \cite{MPFR} or as arbitrary-precision integers with \texttt{GMP} \cite{GMP}. These data types are dynamically allocated and are expensive to allocate and deallocate, which makes FFTs more impractical. Thus, we tailor the data structures to make polynomial multiplication as efficient as possible, since this will be the bottleneck of the algorithm.

We represent each polynomial with a data structure containing the indices and coefficients of monomials with non-zero coefficients. Additionally, we store an ordered list of indices by degrees either intrinsically, or with an auxiliary vector. This information helps speed up the multiplication operations, as we now discuss.

The multiplication procedure begins by constructing a hash map $\mathfrak{P}$ that will store the indices corresponding to coefficients of the resulting polynomial. One then performs two nested for-loops that iterate over the monomials in each polynomial ordered by degree. At each step, one multiplies the monomials, uses the hash map $\mathfrak{M}$ to find the corresponding index, and then multiplies the coefficients and modifies $\mathfrak{P}$ to store the information. Iterating over the monomials sorted by degree allows one to break  out of the inner loop early, as soon as the product of the monomials exceeds the specified maximum degree. This escape significantly speeds up the procedure. The last step is to convert the hash map $\mathfrak{P}$ into the data structure of the resulting polynomial. Note that we could have used a different hash map $\mathfrak{M}'$ that maps pairs of indices to another index, but in practice this uses immoderate amounts of memory when dealing with a large number of monomials.

\begin{mdframed}[linewidth=0.5mm]
\textbf{Toy Multiplication example}

Suppose that the generating monomials are $x_0$ and $x_1$, the grading vector is $(1,1)$, and the maximum degree is 2. The list of sorted monomials and hash map $\mathfrak{M}$ are given by
\begin{equation}
    \mathfrak{L}=\begin{blockarray}{ccc}
& x_0 & x_1 \\
\begin{block}{c(cc)}
  0 & 0 & 0 \\
  1 & 0 & 1 \\
  2 & 1 & 0 \\
  3 & 0 & 2 \\
  4 & 1 & 1 \\
  5 & 2 & 0\\
\end{block}
\end{blockarray}\ \ , \qquad\qquad
\mathfrak{M}=\begin{Bmatrix}
    (0,0) \rightarrow 0\\
    (0,1) \rightarrow 1\\
    (1,0) \rightarrow 2\\
    (0,2) \rightarrow 3\\
    (1,1) \rightarrow 4\\
    (2,0) \rightarrow 5\\
\end{Bmatrix}\ .
\end{equation}
Let us work through the multiplication of $p=2x_1+4x_0x_1$ and $q=3+x_0+3x_1^2$. The first step is to construct an empty hash map $\mathfrak{P}=\{\}$. Then with nested for-loops we multiply the monomials. We start with the product of $2x_1$ and $3$. First we compute $(0,1)+(0,0)$ and use $\mathfrak{M}$ to figure out that the resulting monomial has index $1$. We then multiply the coefficients to obtain $6$. Since $\mathfrak{P}$ does not contain the index $1$ we add the entry $1\rightarrow 6$. Similarly, we then multiply $2x_1$ and $x_0$, and add the entry $4\rightarrow 2$. Finally, before multiplying $2x_1$ by $3x_1^2$ we note that the resulting monomial will be of degree 3, so we skip the computation. We now turn to the next iteration of the outer loop. As before, when multiplying $4x_0x_1$ by $3$ we find that the resulting monomial has index $4$, but now this index is already in $\mathfrak{P}$. Hence, we modify the entry in the hash map to read $4\rightarrow 14$. Finally, we note that multiplying $4x_0x_1$ and $x_0$ would exceed the maximum degree, so we break out of the inner loop. Note that we saved some time by not even considering the product of $4x_0x_1$ and $3x_1^2$, and in more complex examples this kind of time savings is very significant. This illustrates the importance of storing the monomials of a polynomial sorted by degree. The last step of the computation would be to turn the hash map $\mathfrak{P}=\{4\rightarrow 14, 1\rightarrow 6\}$ into the appropriate data structure representing the polynomial $pq=6x_1+14x_0x_1$.
\end{mdframed}

Let us now compute enumerative invariants by carrying out the procedure explained in \S\ref{subsec:enumerative_invariants} and \S\ref{sec:truncation}.
We start by computing the coefficients 
$c_{\vec{n}}$,
$c^a_{\Vec{n}}$, and
$\hat{c}^{ab}_{\Vec{n}}$
in \eqref{eq:fundamental_period_final}, 
\eqref{eq:cadef},
and
\eqref{eq:chat}, respectively, using the formulas presented in \S\ref{sec:Appendix_cformulas}.
Since each coefficient is independent, this can easily be done in parallel.  

These coefficients are then used to construct the 
right-hand side of \eqref{eq:perturbative_GV_formula}, which captures instanton contributions to the derivative of $\mathcal{F}$:
\begin{equation}
\label{eq:repeatperturbative_GV_formula}
    -(2\pi i)^2\mathcal{F}_a^{\text{inst.}} =\phantom{\Biggl(}\frac{1}{2}\kappa_{abc}\frac{\hat{c}^{bc}(\psi)-c^b(\psi)c^c(\psi)}{\varpi_0(\psi)}\, .
\end{equation}
The remaining task is to express the right-hand side as a sum of either $q^{\vec{n}}$ or $\text{Li}_2(q^{\vec{n}})$.
\begin{equation}
\label{eq:rerepeatperturbative_GV_formula}
    -(2\pi i)^2\mathcal{F}_a^{\text{inst.}} =\sum_{\vec{n}\in\mathcal{M}_V}n_a\text{GW}^0_{\vec{n}}\,q^{\Vec{n}} = \sum_{\vec{n}\in\mathcal{M}_V}n_a\text{GV}^0_{\vec{n}}\,\text{Li}_2(q^{\Vec{n}})\, , 
\end{equation}
where 
\begin{equation}
q^{\Vec{n}} := \psi^{\Vec{n}}\,\mathrm{exp} \Bigl({ n_b \tfrac{c^b(\psi)}{\varpi_0(\psi)}}\Bigr)\,.
\end{equation}
This could be done by inverting the series to find $\psi^{\vec{n}}$ as a series in $q^{\vec{n}}$. However, it is much more efficient to extract the enumerative invariants degree-by-degree. The procedure works as follows.

\vspace{3mm}
\setlength{\algomargin}{0em}
\begin{algorithm}[H]
    \begin{enumerate}
        \item Fix a degree $d$, starting from $d=1$.
        \item Find all curves at degree $d$, and form the set $\mathcal{S}:=\{\mathcal{C}\in \mathcal{M}_X|\text{deg}(\mathcal{C})=d\}$. Since at linear order $\psi^{\vec{n}}=q^{\vec{n}}$, the GW or GV invariants can be read off from the coefficient of the corresponding monomial in $\mathcal{F}_a^{\text{inst.}}$, cf.~\eqref{eq:rerepeatperturbative_GV_formula}.  Thus, we extract all invariants from the curves in $S$.
        \item Compute either $q^{\vec{n}}$ or $\text{Li}_2(q^{\vec{n}})$ for all curves in $\mathcal{S}$ in parallel. This is the most expensive step in the entire algorithm, as it involves many polynomial multiplications. To improve efficiency, we keep some intermediate results, so that subsequent computations require fewer polynomial multiplications.
        \item Subtract the computed terms, multiplied by the appropriate factor, so as to eliminate the corresponding monomials in $\mathcal{F}_a^{\text{inst.}}$.
        \item Increment $d$ by one, and repeat from the first step, until the maximum degree is reached.
    \end{enumerate}
    \caption{Find enumerative invariants iteratively.}
\end{algorithm}
\vspace{3mm}

\section{GV Invariants of Toric Curves}\label{sec:toric}

In this section we will obtain formulas
for the GV invariants of very simple curve classes, without using mirror symmetry. 
We will use these analytical results as a check of the correctness of the computation  of GV invariants using mirror symmetry in \S\ref{sec:hktyandbeyond}.  As in \cite{Katz:2022yyg}, we will restrict to rational curves with smooth moduli spaces $\mathcal{M}$ for which the GV invariant is computed by the simple formula
\begin{equation}\label{eq:enumerative_GV_formula}
    \text{GV}^0=(-1)^{\text{dim}(\mathcal{M})}\chi(\mathcal{M})\, ,
\end{equation}
where $\chi$ is the topological Euler characteristic.

\subsection{Complete intersection curves}

First, we will define simple classes of complete intersection curves in Calabi-Yau threefold hypersurfaces, whose GV invariants we will be after.

As in \S\ref{sec:batyrev}, we consider a generic Calabi-Yau hypersurface $X$ in a toric variety $V$ whose toric fan $\Sigma$ is defined via an FRST $\mathcal{T}$ of a reflexive polytope $\Delta^\circ$. 
In this setting, we consider any pair $(p_I,p_J)$ of points in $\Delta^\circ$ associated with toric divisors $(\hat{D}_I,\hat{D}_J)$. 
We then examine the effective curve $\mathcal{C}_{IJ}$ in $X$ that arises via the complete intersection of this pair of toric divisors with the Calabi-Yau hypersurface, i.e.
\begin{equation}
    \mathcal{C}_{IJ}:=\hat{D}_I\cap \hat{D}_J\cap X\equiv D_I\cap D_J\, ,
\end{equation}
with $D_I:=\hat{D}_I\cap X$.  We will focus on the irreducible components of such a curve.

If the line $\ell_{IJ}$ running from $p_I$ to $p_J$ is not an edge of $\mathcal{T}$ then $\hat{D}_I\cap \hat{D}_J=\emptyset$, and thus also $\mathcal{C}_{IJ}=\emptyset$. One draws the same conclusion if $\ell_{IJ}$ is an edge of $\mathcal{T}$ that is not contained in any two-face of $\Delta^\circ$: setting $x_I=x_J=0$ implies that the generic anti-canonical polynomial $f$ degenerates to a monomial, associated with the vertex of $\Delta$ that is dual to the three-face $\Theta_3^\circ \subset \Delta^{\circ}$ containing $\ell_{IJ}$. This monomial does not depend on any coordinates contained in $\Theta_3^\circ$, and thus the requirement $f|_{x_I=x_J=0}=0$ implies that $x_K=0$ for some $K$ associated to a point $p_K$ not contained in $\Theta_3^\circ$. But then the $\{p_I,p_J,p_K\}$ could not possibly lie in the same cone of the toric fan defined by $\mathcal{T}$, and thus again $\mathcal{C}_{IJ}=\emptyset$.

Thus, the only non-trivial complete intersection curves $\mathcal{C}_{IJ}$ arise from pairs of points contained in a shared two-face of $\Delta^\circ$. We will make the following distinction. If the line $\ell_{IJ}$ is contained in a one-face we will say that $\mathcal{C}_{IJ}$ is a \emph{one-face curve}, while if $\ell_{IJ}$ is contained in a two-face but not in any one-face we will say that $\mathcal{C}_{IJ}$ is a \emph{two-face curve}. Similarly, we will call a divisor $D_I$ a \emph{vertex divisor} if it arises from a vertex of $\Delta^\circ$, a \emph{one-face divisor} if it arises from a point interior to a one-face, and a \emph{two-face divisor} if it arises from a point interior to a two-face.

\subsection{Gluing non-compact toric calabi-yau threefolds}

It will turn out to be useful that certain open patches in compact Calabi-Yau threefold hypersurfaces are isomorphic to non-compact Calabi-Yau threefolds that are toric varieties themselves.

Let $\Theta^\circ_2$ be a two-face of $\Delta^\circ$. One may consider a dense open patch $\mathcal{U}(\Theta^\circ_2)\subset V$ defined by requiring that $x_I\neq 0$ for all $I$ such that $p_I\notin \Theta^\circ_2$. This patch $\mathcal{U}(\Theta^\circ_2)$ is itself a toric fourfold $\hat{V}_{\Theta^\circ_2}$, and its toric fan $\hat{\Sigma}_{\Theta_2^\circ}$ arises from the subset of cones in the toric fan of $V$ that are contained in the cone over $\Theta^\circ_2$.
As all cones in $\hat{\Sigma}_{\Theta_2^\circ}$ are contained in a three-dimensional subspace of $N_{\mathbb{R}}$ spanned by the points in 
$\Theta^\circ_2$, we can view $\hat{\Sigma}_{\Theta_2^\circ}$ as a set of cones in this three-dimensional subspace,
defining a 
a toric fan 
$\Sigma_{\Theta_2^\circ}$ 
that in turn defines a toric threefold $V_{\Theta^\circ_2}$. 
We have $\hat{V}_{\Theta^\circ_2}\simeq \mathbb{C}^*\times V_{\Theta^\circ_2}$, where the coordinate $Z$ on the $\mathbb{C}^*$ factor can be viewed as the toric coordinate of an arbitrarily chosen point in $\Delta^\circ$ not contained in $\Theta^\circ_2$. As $V_{\Theta^\circ_2}$ is defined by a toric fan whose edges all lie in the same affine two-dimensional plane, we have that $V_{\Theta^\circ_2}$ is Calabi-Yau, i.e.~$c_1(TV_{\Theta^\circ_2})=0$.

As a simple example consider a two-face $\Theta^\circ_2$ with three vertices and a single interior point, as depicted in Figure \ref{fig:P2fan}. In this case the local Calabi-Yau $V_{\Theta^\circ_2}$ is the total space of the line bundle $\mathcal{O}_{\mathbb{P}^2}(-3)$, where the zero section $\simeq \mathbb{P}^2$ corresponds to the divisor $\pi(D_{I})$ associated with the interior point. 

For a generic point in $V_{\Theta^\circ_2}$, the defining polynomial equation of $X$ amounts to an equation for the $\mathbb{C}^*$ coordinate in $\hat{V}_{\Theta^\circ_2}\simeq \mathbb{C}^*\times V_{\Theta^\circ_2}$, with $g+1$ distinct solution branches, for some $g>0$. Thus we may view the induced patch $\mathcal{U}(\Theta_2^\circ)\cap X$ in $X$ as the branched cover of $g+1$ copies of the toric Calabi-Yau threefold $V_{\Theta^\circ_2}$, with $n$ branches meeting over codimension-$n$ loci in $V_{\Theta^\circ_2}$, 
\begin{equation}\label{eq:projection_map}
    \pi\, : \mathcal{U}(\Theta_2^\circ)\cap X\rightarrow V_{\Theta^\circ_2}\, .
\end{equation}
In other words we can view $\mathcal{U}(\Theta_2^\circ)\cap X$ as a fibration of a set of $g+1$ points over $V_{\Theta^\circ_2}$, and \eqref{eq:projection_map} is the projection to the base.

The number of branches $g$ can be computed by restricting to a toric point in $V_{\Theta^\circ_2}$, i.e. by setting $x_I=x_J=x_K=0$ for $(p_I,p_J,p_K)$ spanning a cone in $\Sigma_{\Theta^\circ_2}$. Then the generic anti-canonical polynomial reduces to a generic linear combination of monomials associated with points in the dual one-face $\Theta_1\subset \Delta$, and only depends on the single $\mathbb{C}^*$ coordinate $Z$: it is independent of the coordinates of $V_{\Theta^\circ_2}$. Letting $k$ be the number of points in $\Theta_1\cap M$, it follows that $g+1=k-1$.  Thus, $g$ is equal to the number of interior points of the dual one-face $\Theta$, i.e.~$g$ is the \emph{genus} of $\Theta^\circ_2$.

\subsection{Irreducible components of complete intersection curves}

In order to compute GV invariants using \eqref{eq:enumerative_GV_formula} we first need to extract the irreducible components of complete intersection curves.

Let $\mathcal{C}_{IJ}$ be a two-face curve associated with a two-face $\Theta^\circ_2$ of genus $g$. Along $\mathcal{C}_{IJ}$ the generic anti-canonical polynomial of $V$ again reduces to a degree $g+2$ polynomial in the $\mathbb{C}^*$ coordinate $Z$, but does not depend on any of the coordinates of $V_{\Theta_2^\circ}$. 
Thus, $\mathcal{C}_{IJ}$ is isomorphic to the disjoint union of $g+1$ copies of $\pi(\mathcal{C}_{IJ})\subset V_{\Theta_2^\circ}$,
\begin{equation}
    \mathcal{C}_{IJ}=\coprod_{i=1}^{g+1}\mathcal{C}^{(i)}_{IJ}\, , 
\end{equation}
where we label each component by an index $i=1,\ldots,g+1$, $\mathcal{C}_{IJ}^{(i)}\simeq \pi(\mathcal{C}_{IJ})$. 

Similarly, each two-face divisor $D_I$ is a disjoint union of $g+1$ copies $D_I^{(i)}\simeq \pi(D_I)$, and each copy contributes a linearly independent class to $H^2(X,\mathbb{Z})$. If there exists at least one two-face divisor $D_I$ with $g>0$ then $h^{1,1}(X)>h^{1,1}(V)$ and $H^2(X,\mathbb{Z})$ is generated by all irreducible components of the $D_I$ \cite{Batyrev:1993oya}. We will not assume that $h^{1,1}(X)=h^{1,1}(V)$ in what follows.  

If $\mathcal{C}_{IJ}$ is such that either $D_I$ or $D_J$ is a two-face divisor, then any copies $\mathcal{C}^{(i)}_{IJ}$ take values in distinct numerical classes, because each copy $\mathcal{C}^{(i)}_{IJ}$ intersects only with the local copy $D_J^{(i)}$ but not with other copies $D_J^{(j)}$ with $j\neq i$ \cite{Braun:2017nhi}. If instead neither $D_I$ nor $D_J$ are two-face divisors then all copies $\mathcal{C}^{(i)}_{IJ}$ have the same numerical class,
\begin{equation}
    [\mathcal{C}^{(i)}_{IJ}]=\frac{[\mathcal{C}_{IJ}]}{g+1}\, .
\end{equation}
Furthermore, we note that the $\mathcal{C}^{(i)}_{IJ}$ are toric varieties themselves, so they are rational curves, i.e.~they are isomorphic to $\mathbb{P}^1$. 
It is typically straightforward to find a triangulation
for which 
any one of the $[\mathcal{C}^{(i)}_{IJ}]$ becomes a generator of the Mori cone of $V_{\Theta^\circ_2}$, 
and each of the $[\mathcal{C}^{(i)}_{IJ}]$ is also a generator of $\mathcal{M}_X$.
In any event, 
whenever shrinking a curve class $[\mathcal{C}^{(i)}_{IJ}]$ in $V_{\Theta^\circ_2}$ induces singularities merely along compact loci in $V_{\Theta^\circ_2}$, we may compute the moduli space of $[\mathcal{C}^{(i)}_{IJ}]$ using the local model $V_{\Theta^\circ_2}$.
 
We wish to compute the GV invariants of  irreducible components 
$[\mathcal{C}^{(i)}_{IJ}]\in H_2(X,\mathbb{Z})$ of two-face 
curve classes $[\mathcal{C}_{IJ}]$.  To this end, we first select a representative $\mathcal{C}^{(i)}_{IJ}$ for some fixed $i$, and consider its moduli space $\mathcal{M}_{[\mathcal{C}^{(i)}]|V_{\Theta_2^\circ}}$, viewed as a curve in $V_{\Theta_2^\circ}$.

\subsection{Moduli space of curves and GV invariants}

We are now ready to construct moduli spaces of curves, and from these, using \eqref{eq:enumerative_GV_formula}, compute GV invariants.

First, from the triple intersection numbers of $X$ we can compute the normal bundle,
\begin{equation}\label{eq:normal_bundle}
    \mathcal{N}_{\mathcal{C}^{(i)}|X}=\mathcal{O}_X(\pi(D_I))|_{\mathcal{C}^{(i)}}\oplus \mathcal{O}_X(\pi(D_J))|_{\mathcal{C}^{(i)}}\simeq \mathcal{O}_{\mathbb{P}^1}(m)\oplus \mathcal{O}_{\mathbb{P}^1}(n)\, ,
\end{equation}
where 
\begin{equation}\label{eq:mndef}
m:=\tfrac{\kappa_{IIJ}}{g+1} \qquad \text{and} \qquad n:=\tfrac{\kappa_{IJJ}}{g+1}\,.
\end{equation}
As we have
 \begin{equation}
     2=\chi(\mathbb{P}^1)=\chi(\mathcal{C}^{(i)})=-\tfrac{\kappa_{IIJ}+\kappa_{IJJ}}{g+1}=-m-n\, ,
 \end{equation}
 one or both of the pair $(m,n)=(m,-2-m)$ must be negative. 
 
 Without loss of generality we set $m\geq n$, and in particular $n<0$. The normal bundle factor corresponding to infinitesimal perturbations of $\mathcal{C}^{(i)}_{IJ}$ away from the divisor $\pi(D_J)$ has no global sections, and thus every point in the connected component of $\mathcal{M}_{[\mathcal{C}^{(i)}_{IJ}]|V_{\Theta_2^\circ}}$ that contains  $\mathcal{C}^{(i)}_{IJ}$ corresponds to a curve contained in $\pi(D_J)$. 
 
\begin{figure}
     \centering
     \begin{tikzpicture}
     \node at (0,0) 
        {\includegraphics[width=16cm,keepaspectratio,clip,trim=2cm 11cm 2cm 2.5cm]{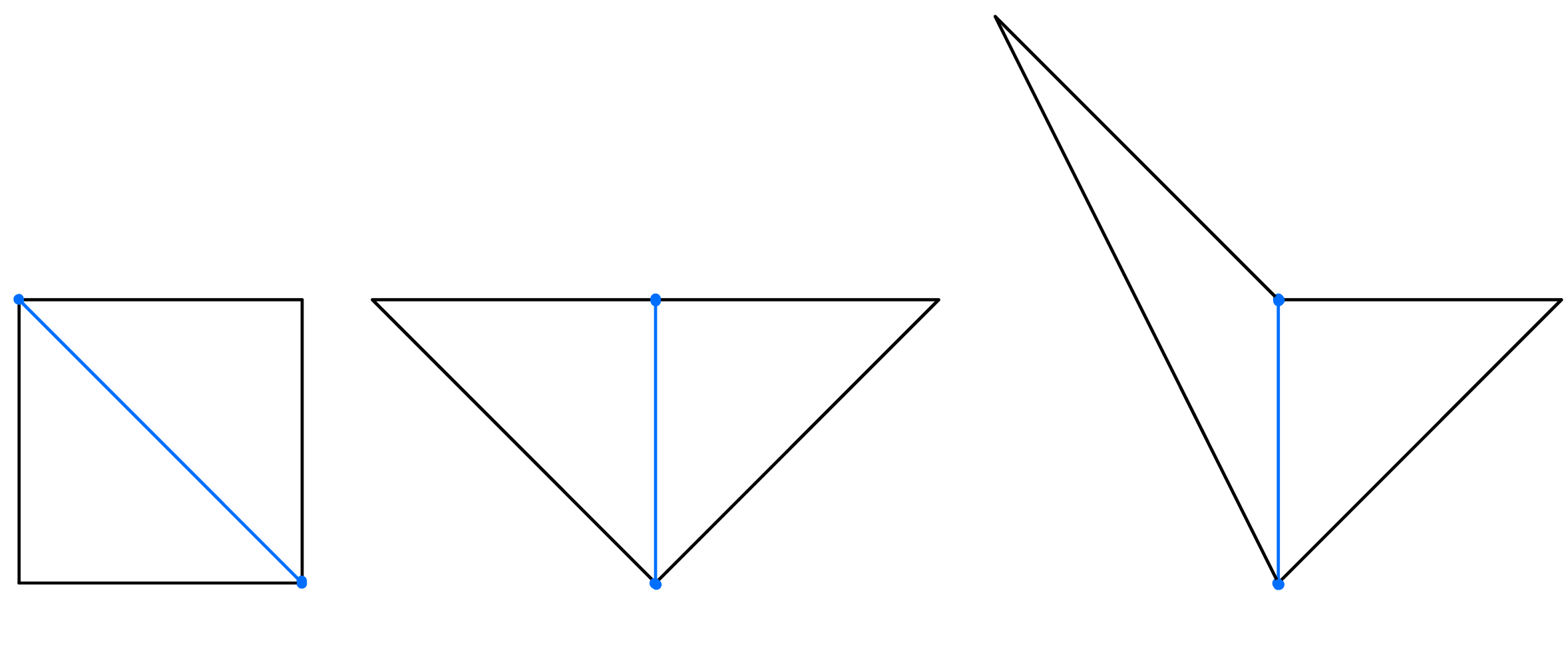}};
     \node at (-8,0.5) 
        {$p_J$};
     \node at (-4.7,-2.9) 
        {$p_I$};
     \node at (-1.2,0.5) 
        {$p_J$};
     \node at (-1.2,-2.9) 
        {$p_I$};
     \node at (5.3,0.5) 
        {$p_J$};
     \node at (5.3,-2.9) 
        {$p_I$};
     \end{tikzpicture}
     \caption{We show the three different relevant classes of two-face curves $\mathcal{C}_{IJ}\simeq \mathbb{P}^1$ arising from edges in the triangulation of a two-face. We depict the edge corresponding to the two-face curve $\mathcal{C}_{IJ}$ in blue, and the adjacent triangles in black. Left: the normal bundle is $\mathcal{O}(-1)\oplus \mathcal{O}(-1)$. Middle: normal bundle $\mathcal{O}\oplus \mathcal{O}(-2)$. Right: normal bundle $\mathcal{O}(1)\oplus \mathcal{O}(-3)$.}
     \label{fig:two_face_curves}
 \end{figure}
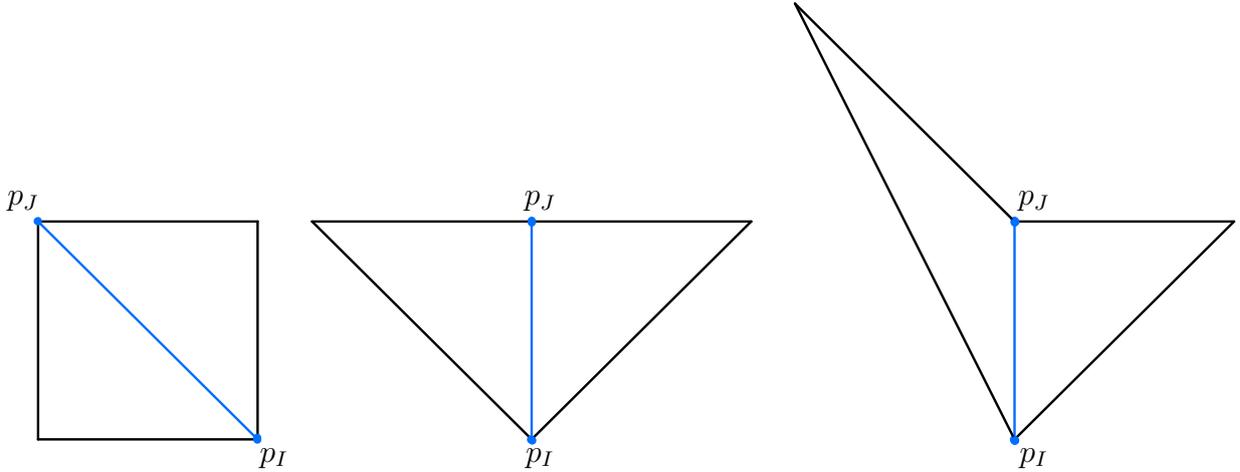

 We now distinguish among three qualitatively different configurations of the triangulation of $\Theta^\circ_2$ around $\ell_{IJ}$: see Figure \ref{fig:two_face_curves}.
 
 \begin{enumerate}
 	\item $(m,n)=(-1,-1)$. In this case the curve $\mathcal{C}^{(i)}_{IJ}$ is an isolated rigid $\mathbb{P}^1$ that shrinks to a conifold singularity, which is clearly compact in $V_{\Theta^\circ_2}$.
 	\item $(m,n)=(0,-2)$: here the curve $\mathcal{C}^{(i)}_{IJ}$ shrinks to a curve worth of $A_1$ singularities in $V_{\Theta^\circ_2}$. The divisor $\pi(D_J)$ degenerates to this curve of singularities, and so the singular locus is compact in $V_{\Theta^\circ_2}$ if and only if $D_J$ is a two-face divisor. 
 	\item $m>0$: here the shrinking divisor $\pi(D_J)$ is compact in $V_{\Theta^\circ_2}$ because convexity of $\Theta^\circ_2$ implies that $D_J$ is a two-face divisor --- see Figure \ref{fig:two_face_curves}.
 \end{enumerate} 
 For now, we restrict to two-face curves $\mathcal{C}^{(i)}_{IJ}$ that yield compact singular loci: that is, either $m \neq 0$, or $m=0$ and $D_J$ is a two-face divisor. We will later return to the case $m=0$ with $D_J$ a one-face divisor.

The normal bundle factor corresponding to infinitesimal perturbations of $\mathcal{C}^{(i)}_{IJ}$ away from the divisor $\pi(D_J)$ has no global sections, and every point in the moduli space $\mathcal{M}_{[\mathcal{C}^{(i)}_{IJ}]|V_{\Theta_2^\circ}}$ corresponds to a curve contained in $\pi(D_J)$. We may therefore equate $\mathcal{M}_{[\mathcal{C}^{(i)}_{IJ}]|V_{\Theta_2^\circ}}$ with the moduli space of curves in $\pi(D_J)$ in the class $[\pi(D_I)|_{\pi(D_J)}]\in H^2(\pi(D_J),\mathbb{Z})$. Our (temporary) simplifying assumption that the singular locus is compact implies that if this moduli space is not just a point, then $\pi(D_J)$ is a toric surface. For a toric surface $\pi(D_J)$  the moduli space $\mathcal{M}_{[\mathcal{C}^{(i)}_{IJ}]|V_{\Theta_2^\circ}}$
is equal to the projectivization of the vector space of global sections $\mathbb{P}\left(\Gamma(\mathcal{L}(\pi(D_I)|_{\pi(D_J)}))\right)$ of the corresponding line bundle $\mathcal{L}(\pi(D_I)|_{\pi(D_J)})$, i.e.
 \begin{equation}
     \mathcal{M}_{[\mathcal{C}^{(i)}]|V_{\Theta_2^\circ}}\simeq \mathbb{P}^{m+1}\, ,
 \end{equation}
 which, in particular, is smooth.
 
 Applying \eqref{eq:enumerative_GV_formula}, the contribution to the GV invariant is
 \begin{equation}\label{eq:toric_GV1}
     (-1)^{\text{dim}(\mathbb{P}^{m+1})}\chi(\mathbb{P}^{m+1})=(-1)^{m+1} (m+2)\, .
 \end{equation}
 If either $D_I$ or $D_J$ is a two-face divisor, the classes $[\mathcal{C}^{(i)}_{IJ}]$ are distinct and the GV invariant is fully accounted for by \eqref{eq:toric_GV1}. Otherwise, the $[\mathcal{C}^{(i)}_{IJ}]$ are all in the same class and so the GV invariant of $[\mathcal{C}^{(i)}]$ picks up a factor $g+1$. In this case, the normal bundle is always $(m,n)=(-1,-1)$. Thus in total we get
 \begin{equation}\label{eq:toric_GV2}
     \text{GV}^0_{[\mathcal{C}^{(i)}_{IJ}]}=\begin{cases}
     (-1)^{m+1} (m+2) & \text{if either $D_I$ or $D_J$ is a two-face divisor\,,}\\
     g+1 & \text{otherwise}\, ,
     \end{cases}
 \end{equation}
where $g$ is the genus of the two-face $\Theta^{\circ}_2$.
We reemphasize that we have assumed that whenever $m=0$ the divisor $D_J$ is a two-face divisor.

 Next, we consider separately the case we have excluded in the above: we let $\mathcal{C}^{(i)}_{IJ}$ be a two-face curve with $m=0$ and with $D_J$ a one-face divisor. We again seek to compute the moduli space of curves in the class $[\mathcal{C}^{(i)}_{IJ}]$. To this end,  we recall from \cite{Braun:2017nhi} that one-face divisors $D_J$ are isomorphic to $\mathbb{P}^1$ fibrations over genus-$g'$ Riemann surfaces $\mathcal{R}_{g'}$, or blow-ups thereof, and each component $\mathcal{C}^{(i)}_{IJ}$ is the fiber of such a fibration,
\begin{equation}
    \mathcal{C}^{(i)}_{IJ}\simeq \mathbb{P}^1\hookrightarrow D_J \twoheadrightarrow \mathcal{R}_{g'}\, ,
\end{equation}
where $g'$ is the genus of the corresponding one-face, i.e.~the number of interior points in the dual two-face of $\Delta$. The moduli space of $\mathcal{C}^{(i)}_{IJ}$ inside $D_J$ corresponds to moving the fiber $\mathbb{P}^1$ across the base, and therefore
\begin{equation}
    \mathcal{M}_{[\mathcal{C}^{(i)}]|X}\simeq \mathcal{R}_{g'}\, ,
\end{equation}
and we compute
\begin{equation}\label{eq:toric_GV_last}
    \text{GV}^0_{[\mathcal{C}_{IJ}^{(i)}]}=2g'-2\, .
\end{equation}

\subsection{Summary}
Let us summarize the results we have assembled for two-face curves.
If $\mathcal{C}_{IJ} = D_I \cap D_J$ is a two-face curve in a two-face 
$\Theta^{\circ}_2$ of genus $g$, $\mathcal{C}_{IJ}$ is the disjoint union of $g+1$ components $\mathcal{C}^{(i)}_{IJ}$, $i=1,\ldots, g+1$, each isomorphic to a $\mathbb{P}^1$, and these components fall into distinct numerical classes if and only if $D_I$ or $D_J$ is a two-face divisor.
We have given formulas for the genus-zero GV invariants of all such curves.
In terms of $m$ defined in \eqref{eq:mndef}, we obtained \eqref{eq:toric_GV2}, which applies whenever $m \neq 0$, and also applies if $m=0$ and $D_J$ is a two-face divisor.  In the remaining case that $m=0$ and $D_J$ is not a two-face divisor, we have instead \eqref{eq:toric_GV_last}.

As an obvious extension of the above one can study moduli spaces of rational curves associated with complete intersections $\mathcal{C}_{I,D}^{(i)}:=\hat{D}_I^{(i)}\cap D\cap X$ where $D$ is a general divisor, with $D_I^2 D<0$. In this case the moduli space of $\mathcal{C}_{I,D}^{(i)}$ is again equivalent to the moduli space of divisors in the class $[D]_{D_I^{(i)}}\in H^2(D_I^{(i)},\mathbb{Z})$. If $D_I^{(i)}$ is toric, the GV invariant is again given by \eqref{eq:toric_GV1} with
\begin{equation}
    m = D_I^{(i)} D^2\, .
\end{equation}
We carry out this computation in the case of the mirror quintic in \S\ref{sec:mirrorquintic}.
Alternatively, one can 
employ the topological vertex formalism \cite{Aganagic:2003db} to compute the GV invariants of general compact curves in $V_{\Theta^\circ_2}$, in order to predict the GV invariants of broader classes of curves in $X$.  Results along these lines appear in \cite{Hayashi:2019fsa} for certain classes of curves in the mirror quintic.

\section{Examples}\label{sec:examples}

In this section we will illustrate our methods in four examples: specifically, in Calabi-Yau threefold hypersurfaces with $(h^{1,1},h^{2,1})=(2,272)$, $(101,1)$, $(251,251)$, and $(491,11)$. 

For any Calabi-Yau threefold obtained from the Kreuzer-Skarke list, we can compute GV invariants directly, in the full-dimensional Mori cone, after making an appropriate choice of grading vector.  Below we give the results of such a full-dimensional computation in each of our examples.

Even so, the number of curve classes below a given degree is exponential in $h^{1,1}$, so storing and manipulating the list of GV invariants becomes expensive at large $h^{1,1}$, which limits one to relatively low degrees.
At the same time, for large $h^{1,1}$ the vast majority of effective curves have vanishing GV invariant.  A more informative approach for $h^{1,1} \gtrsim \mathcal{O}(100)$ is to compute GV invariants only in $p$-faces of the Mori cone, for $p \ll h^{1,1}$.  In the examples below we will take $p=2$, because the resulting invariants are readily displayed in a table.

All Calabi-Yau threefolds considered in this section are generic anti-canonical hypersurfaces of toric varieties defined via an FRST $\mathcal{T}$ of a reflexive polytope $\Delta^\circ$, as in \S\ref{sec:batyrev}.

\subsection{A threefold with $h^{1,1}=2$, $h^{2,1}=272$}

As our first example we consider the threefold with $h^{1,1}=2$, $h^{2,1}=272$ obtained as a hypersurface in $\mathbb{CP}_{[1,1,1,6,9]}$. 
We denote this threefold by $X_{2,272}$.
Because the Mori cone of $X_{2,272}$ is simplicial, there is a canonical grading vector --- the sum of the two extremal rays of the K\"ahler cone --- and a canonical definition of degree.

The enumerative invariants of $X_{2,272}$ are very well known at low degree \cite{Candelas:1994hw}.  We have applied the methods explained in \S\ref{sec:hktyandbeyond} and \S\ref{sec:algorithm} to compute the GV invariants 
of $X_{2,272}$ up to degree 200: see Table \ref{tab:timing}.  This computation takes about six hours.  As an example, for the curve class with degree $(43, 157)$ we find the GV invariant
\medskip

$\text{GV}^{0}_{[43,157]}=$
\seqsplit{
7262452829955349348898970729150141416679110693973320028351068818848643265325138205809181587654546622510792000605295623368284784338445060248524382105750048410452442525195264296260736301841087754901511418496364363866593128124744924629990423053247956813814650497023692754817126168697665924217334360657193483195016680692788147547267353684581662054434046775208949578420733506288158136663478355005217789631576603692419373884907789272751959442276020880914706818542923846400}.

\subsection{Mirror quintic threefold with $h^{1,1}=101$, $h^{2,1}=1$}\label{sec:mirrorquintic}

As our next example we consider the mirror quintic. We have $h^{1,1}=101$ and $h^{2,1} = 1$, and we will denote the mirror quintic by $X_{101,1}$.

The reflexive polytope $\Delta^\circ\subset N$ is defined by vertices $(p_1,\ldots,p_5)$ equal to the columns of
\begin{equation}
    \begin{pmatrix}
        -1 & 4 & -1 &-1 &-1\\
        -1 & -1 & 4 &-1 &-1\\
        -1 & -1 & -1& 4 &-1\\
        -1 & -1 & -1&-1 & 4
    \end{pmatrix}\, .
\end{equation}
For later reference we also define the following points interior to one-faces of $\Delta^\circ$,
\begin{equation}
    \begin{pmatrix}
        p_7 & p_8
    \end{pmatrix}:=\begin{pmatrix}
        -1 & -1\\
        -1 & -1\\
        -1 & -1\\
        1  & 2
    \end{pmatrix}\, ,
\end{equation}
as well as the following points interior to two-faces,
\begin{equation}
    \begin{pmatrix}
        p_{50} & p_{68} & p_{71} & p_{100}
    \end{pmatrix}:=
    \begin{pmatrix}
        -1 & -1 &  0 & 2\\
        -1 &  2 & -1 & -1\\
        1  &  0 & -1 & -1\\
        1  & -1 &  1 & 0
    \end{pmatrix}\, .
\end{equation}

For the Delaunay 
FRST $\mathcal{T}$ of $\Delta^\circ$ we have applied the methods explained in \S\ref{sec:hktyandbeyond} and \S\ref{sec:algorithm} to compute the GV invariants of the mirror quintic.
We take the grading vector to be $[\ell] + [\gamma]$, where $([\ell],[\gamma])\in H^2(X,\mathbb{Z})$ are certain divisor classes considered in \cite{Katz:2022yyg,Hayashi:2019fsa}.
 
Computing GV invariants up to degree 1 involves examining 305 nontrivial curves, all of which turn out to have non-zero GV invariants.  This computation takes a fraction of a second.
Working instead up to degree 4, one examines  360,378,675 curves, of which 2500 have nonzero GV invariants: see Table \ref{tab:timing}.
This computation takes more than a day and requires 240 GB of RAM.
 
\begin{figure}
\centering
\includegraphics[width=6cm]{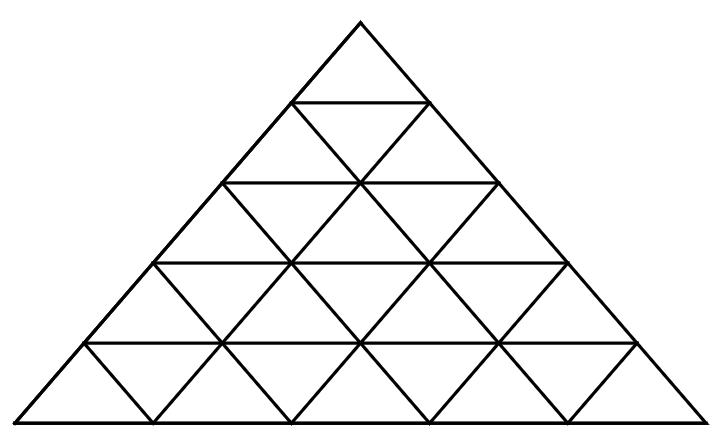}
\caption{The symmetric triangulation of the two-faces of $\Delta^\circ$ for $X_{101,1}$.}\label{fig:2face101}
\end{figure}

As a cross check, we note that certain coarse-grained enumerative invariants of the mirror quintic, corresponding to specific sums of GV invariants,
have been computed in \cite{Hayashi:2019fsa}, for this particularly symmetric FRST of $\Delta^\circ$ (we depict its induced triangulation of the two-faces in Figure \ref{fig:2face101}).\footnote{Furthermore, in the related work \cite{Katz:2022yyg} the GV invariants of certain rational curves were computed directly.}
These are defined as
\begin{equation}
    n^0_{i,j}:=\sum_{[\mathcal{C}]\in \mathcal{M}_X:\, \langle \mathcal{C}, \ell \rangle= i\, ,\, \langle \mathcal{C}, \gamma \rangle= j}\text{GV}^0_{[\mathcal{C}]}\, .
\end{equation}
We have obtained the $n^0_{i,j}$ for $i+j\leq 4$   directly from the GV invariants, and our results agree with those of \cite{Katz:2022yyg,Hayashi:2019fsa}.

Furthermore, we have used the past light cone method introduced in \S\ref{sec:truncation} to compute GV invariants along a two-dimensional face of the Mori cone. In general, even finding extremal rays of the Mori cone is not straightforward, particularly because the Mori cone of the toric ambient variety is typically larger than the Mori cone of the Calabi-Yau hypersurface. In practice, however, one can find low-dimensional faces of the Mori cone via a trial and error method, as follows. Given an irreducible curve class $[\mathcal{C}]$ --- say one inherited from the toric ambient variety via a complete intersection of toric divisors --- one executes the computation of GV invariants using the set of curves
\begin{equation}\label{eq:1dray}
    \mathcal{S}=\{[\mathcal{C}],2[\mathcal{C}],\ldots,n[\mathcal{C}]\}\, ,
\end{equation}
up to some suitably large multiple $n$.

Following the discussion of \S\ref{sec:truncation}, if $[\mathcal{C}]$ is \emph{not} a generator of the Mori cone of the Calabi-Yau hypersurface, this GV computation is inconsistent. Now, from the perspective of the computation detailed in \S\ref{sec:hktyandbeyond} the integrality of GV invariants is due to a miraculous cancellation between rational terms. Therefore, one expects that the above computation will not return integer GV invariants unless $[\mathcal{C}]$ is a generator of the Mori cone, or is not effective at all. Thus, we conjecture that a one-dimensional ray as in \eqref{eq:1dray}, along which the GV computation returns non-trivial integer invariants, corresponds to a generator of the Mori cone, and the resulting integers are the GV invariants along the corresponding one-face of the Mori cone.

Having found a set of generators of the Mori cone, one can consider the two-dimensional cones spanned by pairs of these, and again execute the GV computation along these sub-cones of the Mori cone. Again, if the computation returns integer invariants we conjecture that the two-dimensional sub-cone is a two-face of the Mori cone.  One  could continue to higher-dimensional sub-cones, but we stop at two-dimensional sub-cones because displaying the results remains convenient.

In this way, for an FRST $\mathcal{T}$ obtained with a placing/pushing algorithm with TOPCOM \cite{Rambau2002},
we have found a pair of curve classes $([\mathcal{C}_1],[\mathcal{C}_1])$ that span a two-face of the Mori cone.\footnote{For this purpose the placing/pushing triangulation
has an advantage over the more symmetric
Delaunay triangulation used above.  In asymmetric triangulations, the cone hosting infinite sequences of non-zero GV invariants tends to be near a boundary of the Mori cone, so it is easier to find faces populated by many non-zero GV invariants.} We define the extremal generators via their intersection pairing $q^{[\mathcal{C}]}_I:=\int_X [\mathcal{C}]\wedge [D_I]$ with the prime toric divisors $D_I$. The non-vanishing components are
\begin{align}\label{eq:curve_charges}
    q^{[\mathcal{C}_1]}_{50,68,100}=(1,1,1)\, ,\quad q^{[\mathcal{C}_2]}_{7,   8,  71, 100}=(1,1,-3,1)\, .
\end{align}
The computation of GV invariants along this two-face is highly efficient: we find $\mathcal{O}(10^3)$ non-vanishing GV invariants in $\mathcal{O}(10)$ seconds on a laptop. We give the leading GV invariants in Table \ref{tab:GV_table_mquintic}.
\begin{table}
	\scriptsize \renewcommand{\arraystretch}{0.9}
	\begin{align*}
		\begin{array}{c|cccccccccc}
			\mathdiagbox[width=0.7cm,height=0.5cm,innerleftsep=0.1cm,innerrightsep=0cm]{k}{l\hspace{0.1cm}} & 0 & 1 & 2 & 3 & 4 & 5 & 6 & 7 \\ \hline
			  0          & * & 3 &-6 & 27 & -192&   1695&   -17064 & 188454  \\
            1          & 54 & 18&  -82  &  612 & -5850 &    64478  &     -779058 & 10035288  \\
            2          & 54 & 85& -684  & 7425 &-93320 &  1274238  &  -18353016 & 274391046   \\
            3          & 72 & 312 &-4140& 63846 &-1039290&17415504 &   -297187974 & 5134670850 \\
            4          & 54 & 945 &-20440& 436338 &-9074592& 185055084 &-3719815650 & 73961593398  \\
            5          & 54 &2620 &-87318& 2523096&-66193218&1629543600&-38372051916 & 874495652252 \\
            6          & 72 &6783 &-334128& 12819195&-419524080&12376238193& -339633759600 & 8837783925906\\
            7 & 54 &16200 &-1168632&  58690260 &-2372561334&83309928232&-2651437950876 & 78491030044392 
		\end{array}
	\end{align*}
	\caption{GV invariants $\text{GV}^0_{k[\mathcal{C}_1]+l[\mathcal{C}_2]}$ of $X_{101,1}$ for $l,k\leq 7$.}
	\label{tab:GV_table_mquintic}
\end{table}

Finally, we note that the curve $\mathcal{C}_2$ can be represented by the complete intersection of prime toric divisors $\mathcal{C}_2=D_{7}\cap D_{71}$, and along the facet of the K\"ahler cone where $\mathcal{C}_2$ collapses, the toric divisor $D_{71}$ shrinks to a point.

\subsection{A threefold with $h^{1,1}=h^{2,1}=251$} 

For our next example we consider the reflexive polytope $\Delta^\circ$ whose vertices $(p_1,\ldots,p_5)$ are the columns of 
\begin{equation}
    \begin{pmatrix}
        -903 & 0 & 0 & 0 & 1\\
        -602 & 0 & 0 & 1 & 0\\
        -258 & 0 & 1 & 0 & 0\\
        -42  & 1 & 0 & 0 & 0
    \end{pmatrix}\, .
\end{equation}
Again, we define a further set of points that will become relevant momentarily. The following points are interior to one-faces of $\Delta^\circ$,
\begin{equation}
    \begin{pmatrix}
        p_7 & p_{12} & p_{22} & p_{31} & p_{40}
    \end{pmatrix}:=\begin{pmatrix}
        -861& -774& -602& -451& -301\\
        -574& -516& -401& -301& -200\\
        -246& -221& -172& -129&  -86\\
         -40&  -36&  -28&  -21&  -14
    \end{pmatrix}\, ,
\end{equation}
and we further consider the following points interior to two-faces,
\begin{equation}
    \begin{pmatrix}
        p_{101} & p_{172}
    \end{pmatrix}:=
    \begin{pmatrix}
        -409& -193\\
        -273& -129  \\
        -117&  -55 \\       
         -19&   -9
    \end{pmatrix}\, .
\end{equation}
This polytope is self-dual. A fine, star, regular triangulation of $\Delta^{\circ}$ defines a toric variety in which the generic anti-canonical hypersurface is a smooth Calabi-Yau threefold with $h^{1,1}=h^{2,1}=251$. As before, we choose the Delaunay triangulation, and denote the resulting Calabi-Yau threefold by $X_{251,251}$.

We have applied the methods explained in \S\ref{sec:hktyandbeyond} and \S\ref{sec:algorithm} to compute the GV invariants of $X_{251,251}$.
As a simple choice of grading vector, one can take the tip of the stretched K\"ahler cone, sufficiently scaled and rounded to be integral.  Computing GV invariants up to degree 20 in this grading involves examining 1,047,796 curves, of which 557 have non-zero GV invariants:
see Table \ref{tab:timing}.
This computation takes about two minutes on a desktop computer.  
 
Moreover, as in the previous example, we also compute GV invariants along a (conjectured) two-face of the Mori cone obtained from a placing/pushing triangulation.
The face is generated by the curves $(\mathcal{C}_1,\mathcal{C}_2)$ with non-vanishing intersection pairings
\begin{align}\label{eq:curve_charges2}
    q^{[\mathcal{C}_1]}_{1,  7,  22,  31, 172}=(-3,1,2,1,1)\, ,\quad q^{[\mathcal{C}_2]}_{1,12,31,40,101}=(-3,2,1,1,1)\, .
\end{align}
We depict the GV invariants along the two-face in Table \ref{tab:GV_table_251}.
\begin{table}
	\scriptsize \renewcommand{\arraystretch}{0.9}
	\begin{align*}
		\begin{array}{c|cccccccccc}
			\mathdiagbox[width=0.7cm,height=0.5cm,innerleftsep=0.1cm,innerrightsep=0cm]{k}{l\hspace{0.1cm}} & 0 & 1 & 2 & 3 & 4  \\ \hline
			  0          & * &7& -17& 87& -670\\
            1          & 7 &-750&33940&-1131019&32301672\\
            2          & -17&33940&-7859136&882187785&-66568176839\\
            3          &87&-1131019&882187785&-251747830110&41218265699923\\
            4          &-670&32301672&-66568176839&41218265699923&-12971328454531146\\
            5          &6283&-841123371&3896195467041&-4696379732519631&2614426093633512280\\
            6          &-66432&20600533625&-191037468993574&413491534885669787&-382665656035954784333\\
		\end{array}
	\end{align*}
	\caption{GV invariants $\text{GV}^0_{k[\mathcal{C}_1]+l[\mathcal{C}_2]}$ of $X_{251,251}$ for $l\leq 4$ and $k\leq 6$.}
	\label{tab:GV_table_251}
\end{table}

\subsection{A threefold with $h^{1,1}=491$, $h^{2,1}=11$}
 
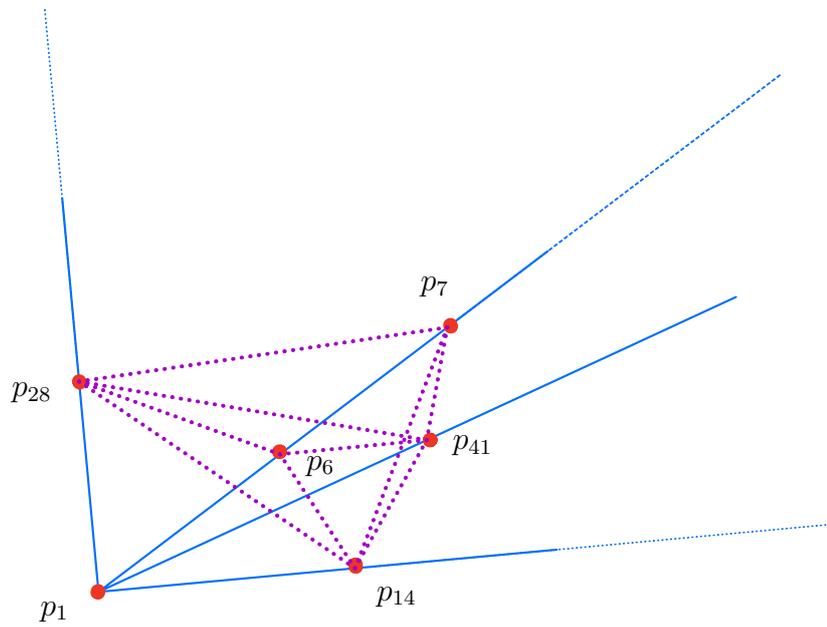
\begin{figure}
     \centering
     \begin{tikzpicture}
     \node at (0,0) 
        {\includegraphics[width=12cm,keepaspectratio,clip,trim=2cm 3cm 2cm 4cm]{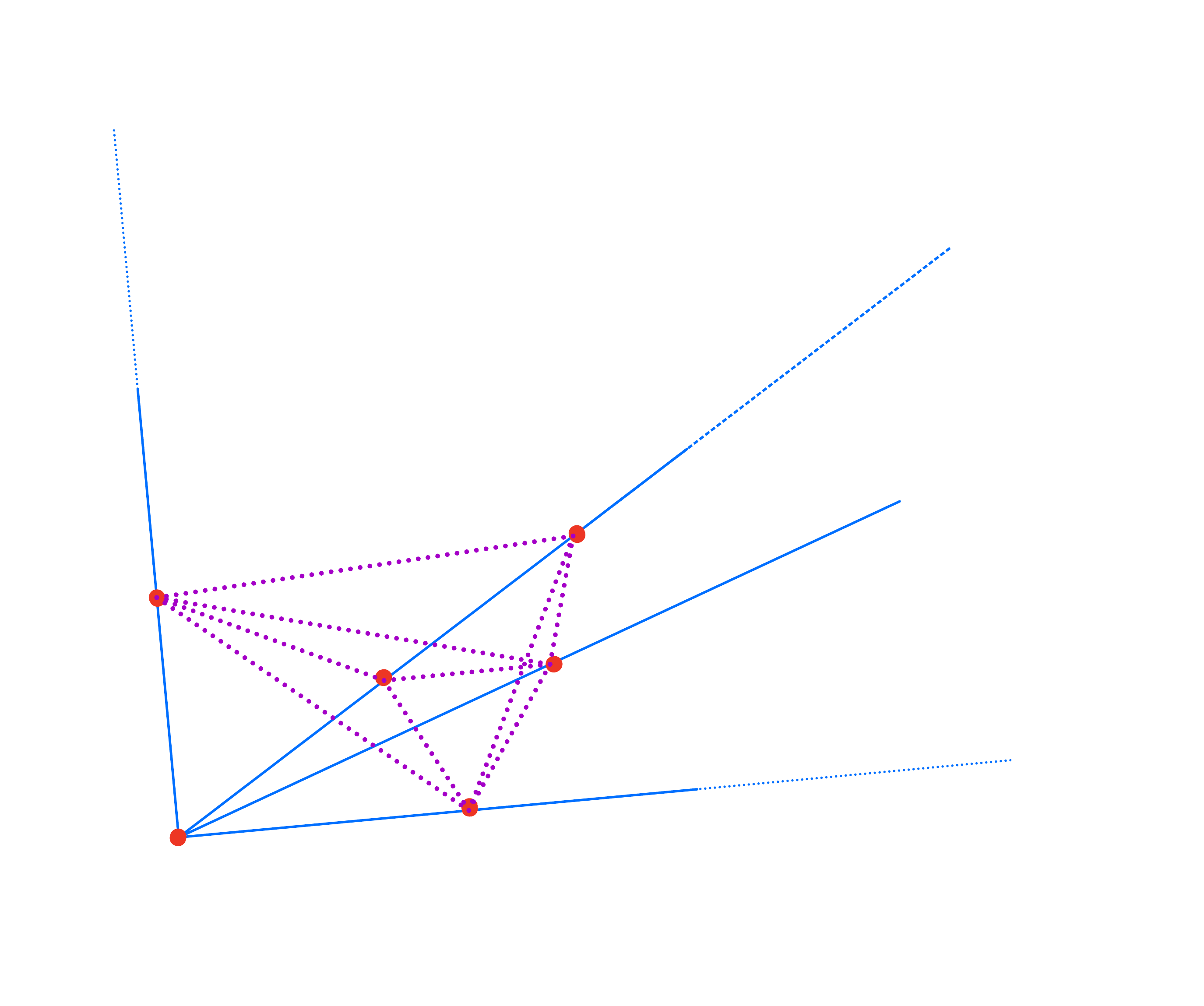}};
     \node at (-5.8,-1) 
        {$p_{28}$};
     \node at (-5.5,-3.9) 
        {$p_1$};
     \node at (-0.5,0.4) 
        {$p_7$};
     \node at (-2.00,-1.97) 
        {$p_6$};
     \node at (0.0,-1.7) 
        {$p_{41}$};
     \node at (-1.0,-3.7) 
        {$p_{14}$};
     \end{tikzpicture}
     \caption{Cartoon of the neighborhood of the vertex $p_1\in \Delta^\circ$. Blue lines represent one-faces ending in $p_1$, and purple dotted lines represent edges of the triangulation of two-faces. Note that this illustration has its shortcomings: the seemingly interior one-face containing $(p_6,p_7)$ is a one-face of three distinct three-faces, and each of these includes two of the remaining one-faces, but not the third.}
     \label{fig:491-triangulation}
 \end{figure}

Our final example begins with the reflexive polytope $\Delta^\circ$ whose vertices $(p_1,\ldots,p_5)$ are the columns of 
\begin{equation}
    \begin{pmatrix}
       -63&   0&   0&   1&  21\\
       -56&   0&   1&   0&  28\\
       -48&   1&   0&   0&  36\\
       -42&   0&   0&   0&  42
    \end{pmatrix}\, .
\end{equation}
We will also need the following points,
\begin{equation}
    \begin{pmatrix}
        p_6 & p_7 & p_{14}& p_{28}&p_{41}
    \end{pmatrix}:=\begin{pmatrix}
       -62& -61& -54& -42& -31\\
       -55& -54& -48& -37& -28\\
       -47& -46& -41& -32& -24\\
       -41& -40& -36& -28& -21
    \end{pmatrix}\, .
\end{equation}
The points $(p_6,p_{14},p_{28},p_{41})$ are the unique points closest to $p_1$ strictly interior to one of the respective four one-faces of $\Delta^\circ$ that have $p_1$ as a vertex. The point $p_7$ is the next point along the one-face containing $p_6$: see Figure \ref{fig:491-triangulation}.

Any FRST of $\Delta^{\circ}$ defines a toric variety in which the generic anti-canonical hypersurface is a smooth Calabi-Yau threefold with $h^{1,1}=491, h^{2,1}=11$.  We note that $491$ is the largest known value of $h^{1,1}$ of a Calabi-Yau threefold. 
We now consider an FRST, and denote the resulting Calabi-Yau threefold by $X_{491,11}$. The only relevant property of our chosen FRST is that the following pairs of points from the above set of points interior to one-faces are connected by an edge of the triangulation,
\begin{equation}
    (6,14)\, , \quad (6,28)\, ,\quad (6,41)\, ,\quad (7,14)\, ,\quad (7,28)\, ,\quad (7,41)\, ,
\end{equation}
as illustrated in Figure \ref{fig:491-triangulation}.

First, we have applied the methods explained in \S\ref{sec:hktyandbeyond} and \S\ref{sec:algorithm} to compute the GV invariants of $X_{491,11}$: see Table \ref{tab:timing}.
For the grading vector, we take the tip of the stretched K\"ahler cone, scaled and rounded to be integral.
Computing to degree 25 in this grading involves examining 1,699,192 curves, of which 627 have nonzero GV invariants.  This process takes about nine minutes on a desktop computer.

Finally, we   consider a two-face of the Mori cone of $X_{491,11}$. This two-face is spanned by the curves $(\mathcal{C}_1,\mathcal{C}_2)$, which are specified by their non-vanishing intersection pairings
\begin{align}\label{eq:curve_charges3}
    q^{[\mathcal{C}_1]}_{1,6,7}=(1, -2,  1)\, ,\quad  q^{[\mathcal{C}_2]}_{1,  6, 14, 28, 41}=(-3,  1,  1,  1,  1)\,  .
\end{align}
Both curves can be represented as complete intersection curves:
\begin{equation}
    \mathcal{C}_1=D_6\cap D_7\, ,\quad \, \mathcal{C}_2=D_1\cap D_6\,  .
\end{equation}
In this instance, the curves $(\mathcal{C}_1,\mathcal{C}_2)$ span a two-face of the Mori cone inherited from the toric ambient variety, so the computation of GV invariants along it is predicted to correctly return the integer GV invariants. We present the results in Table \ref{tab:GV_table_491}.
\begin{table}
	\scriptsize \renewcommand{\arraystretch}{0.9}
	\begin{align*}
		\begin{array}{c|ccccccccccc}
			\mathdiagbox[width=0.7cm,height=0.5cm,innerleftsep=0.1cm,innerrightsep=0cm]{k}{l\hspace{0.1cm}} & 0 & 1 & 2 & 3 & 4 & 5 & 6 & 7 & 8 & 9  \\ \hline
			  0          &   *&3&-6&27&-192&1695&-17064& 188454&-2228160&27748899\\
            1          &  -2&   4&-10& 64& -572&   6076&   -71740&     909760&      -12146622&      168604540\\
            2          &    0&    3& -12&             91&
                 -980&          12259&        -166720&        2394779&
             -35737460&      548460000\\
            3          &   0&5&   -12&          108&               -1332&          18912& -289440&       4632120&           -76306398&     1282295808\\
            4          &  0&         7&            -24&            150&
                -1808&          26983&        -443394&        7665776&
            -136440800&     2471539911\\
            5          &   0&             9&            -56&            294&
 -2982&   42005& -689520& 12254816& -227540162&     4331108122\\
            6          & 0&    11&           -140&            675&  -5992&          76608&      -1192644&      20764870&
            -386343036&     7482057534\\
            7 & 0&             13&           -324&           1738&
      -13550&         158814&       -2322056&       38750866&   -703362386&    13488597425\\
            8 & 0&             15&           -686&           4732&         -33552&         359898&       -4954570&       79050699&        -1387505216&    25992283043\\
            9&0&            17&          -1328&          12960& -88746&   874588&   -11327904&      172924796&      -2932945300&    53475853968\\
		\end{array}
	\end{align*}
	\caption{GV invariants $\text{GV}^0_{k[\mathcal{C}_1]+l[\mathcal{C}_2]}$ of $X_{491,11}$ for $k,l\leq 9$.}
	\label{tab:GV_table_491}
\end{table}
In the limit where $\mathcal{C}_1$ shrinks, the one-face divisor $D_6$ shrinks to a curve of genus zero, leading to non-abelian $\mathfrak{su}(2)$ enhancement \cite{Aspinwall:1995xy,Katz:1996ht}, while in the limit where $\mathcal{C}_1$ shrinks, the vertex divisor $D_1$ shrinks to a point, leading to a tensionless string CFT \cite{Witten:1996qb}.

\begin{table}[] 
\begin{center}
\begin{tabular}{ccccc}
 $h^{1,1}$ & degree & $N_{\text{points}}$    & $N_{\text{nonzero}}$ & Time  \\
 \hline
 2   &  200  & 20,300       &  20,300  &   6 h   \\
 101 &  2    & 46,710      & 560     & 4 s      \\
 101 &  4    & 360,378,675 & 2,500    & 1 d    \\
 251 &  20   & 1,047,796   & 557     & 2 m   \\
 491 &  25   & 1,699,192   & 627     & 9 m   
\end{tabular}
\end{center}
\caption{Size and duration of the computation in the examples, in seconds, minutes, hours, and days. The number of curve classes examined is denoted $N_{\text{points}}$,
while $N_{\text{nonzero}}$ is the number of nonzero GV invariants found.}
\label{tab:timing}
\end{table}

\section{Conclusions}\label{sec:conclusions}

The main result of this work is an efficient algorithm for using mirror symmetry to compute the prepotential in type II compactifications on Calabi-Yau threefold hypersurfaces. 
Specifically, we aimed to be able to compute worldsheet instanton corrections to the prepotential for any threefold arising from the Kreuzer-Skarke list \cite{Kreuzer:2000xy}.

The classic paper of Hosono, Klemm, Theisen, and Yau \cite{HKTY} laid out a procedure that is valid for Calabi-Yau threefold hypersurfaces in which the Mori cone is simplicial.  One computes a fundamental period on the type IIB side, uses properties of the Picard-Fuchs system to obtain the remaining periods, and finally reads off the Gopakumar-Vafa invariants.

The method of \cite{HKTY} has two key limitations.  The first is that the vast majority of threefolds resulting from the Kreuzer-Skarke list --- and in particular, almost all such threefolds with large $h^{1,1}$ --- have non-simplicial Mori cones.
The second is that extracting enumerative invariants at large $h^{1,1}$ appears exponentially costly.

In this work we overcame the above limitations: we devised a generalization of \cite{HKTY} that is valid for any threefold hypersurface, and we produced an implementation that is practical even for $h^{1,1}$ as large as 491.
This improved capability rests on a series of conceptual advances that we presented in \S\ref{sec:hktyandbeyond} and \S\ref{sec:algorithm}.

We illustrated our method in a collection of examples.
In the hypersurface in $\mathbb{CP}_{[1,1,1,6,9]}$, with $h^{1,1}=2$, we computed GV invariants to degree 200.  Even for the largest-known value of $h^{1,1}$, i.e.~491, we were able to obtain hundreds of GV invariants in minutes on a desktop computer: see Table \ref{tab:timing}.

Our results have immediate applications to the study of instanton corrections in Calabi-Yau compactifications, and have already been used in the construction of AdS vacua in \cite{Demirtas:2021nlu}, and of complete K\"ahler moduli spaces in \cite{Gendler:2022ztv}.  We plan to incorporate an implementation of our algorithm in a near-future version of \texttt{CYTools}.

\section*{Acknowledgments}

We thank Mike Douglas, Naomi Gendler, Ben Heidenreich, Richard Nally, Andreas Schachner, Mike Stillman and Tom Rudelius for helpful discussions.
The research of M.D.~was supported in part by the
National Science Foundation under Cooperative Agreement PHY-2019786 (The NSF
AI Institute for Artificial Intelligence and Fundamental Interactions). 
M.K.~was supported by a Pappalardo Fellowship.
L.M., J.M., and A.R.-T.~were supported in part by NSF grant PHY-2014071.

\appendix
\section{Coefficient Formulas}\label{sec:Appendix_cformulas}
In this Appendix, we give some concrete formulas for the coefficients $c_{\Vec{n}}^a$ and $\hat{c}_{\Vec{n}}^{ab}$ that appear in \S\ref{subsec:enumerative_invariants}. We define $k_I:={Q^a}_I n_a$, as well as $k_0:={Q^a}_0 n_a$. 

First, we reproduce from Appendix A of \cite{HKTY} that if $\Vec{n}$ is such that $k_I\geq 0$ for all $I$, then 
\begin{align}
    c_{\Vec{n}}&=\frac{k_0!}{\prod_{I} k_I !}\, ,\\
    c^a_{\Vec{n}}&=\mathcal{A}^a_{\vec{n}} \, c_{\Vec{n}}:=\Bigl[{Q^a}_0 \Psi(1+k_0)-\sum_I {Q^a}_I \Psi(1+k_I)\Bigr] c_{\Vec{n}}\, ,\\
    c^{ab}_{\Vec{n}}&=\left(\mathcal{A}_{\vec{n}}^a \mathcal{A}_{\vec{n}}^b +\mathcal{B}^{ab}_{\Vec{n}}\right)c_{\vec{n}}\, ,\\ 
    \text{with}\quad \mathcal{B}^{ab}_{\Vec{n}}&:={Q^a}_0{Q^b}_0\Psi'(1+k_0)-\sum_I {Q^a}_I {Q^b}_I \Psi'(1+k_I)\, ,
\end{align}
where $\Psi(z):=\del_z \log \Gamma(z)\equiv \Psi^{(1)}(z)$, and $\Psi'(z)\equiv \Psi^{(2)}(z)$, where $\Psi^{(n)}(z)$ denotes the polygamma function of order $n$. 
Using the identities
\begin{equation}
    \Psi(1+m)=H_m-\gamma_E\, ,\quad \Psi'(1+m)=-H_{m,2}+\frac{\pi^2}{6}\, ,\quad \forall m\in \mathbb{N}\, ,
\end{equation}
where $H_{m,n}\in \mathbb{Q}$ are the generalized harmonic numbers, $H_m\equiv H_{m,1}$, and $\gamma_E$  is the Euler–Mascheroni constant, we obtain\footnote{We thank Andreas Schachner for a useful observation relating to these expressions that led us to improve the implementation of our algorithm.}
\begin{align}\label{eq:arat}
    \mathcal{A}_{\Vec{n}}^a&={Q^a}_0 H_{k_0}-\sum_I {Q^a}_I H_{k_I}\, ,\\  \label{eq:brat}
    \mathcal{B}^{ab}_{\Vec{n}}&=\hat{\mathcal{B}}_{\Vec{n}}^{ab}+\frac{\pi^2}{6}\Bigl[{Q^a}_0{Q^b}_0-\sum_I {Q^a}_I{Q^b}_I\Bigr]\, ,\\ \label{eq:crat}
    \text{with}\quad \hat{\mathcal{B}}_{\Vec{n}}^{ab}&:=\sum_I {Q^a}_I {Q^b}_I H_{k_I,2}-{Q^a}_0 {Q^b}_0 H_{k_0,2}\, .
\end{align}
The vanishing of the first Chern class, $c_1(X)=0$, corresponds to the condition ${Q^a}_0-\sum_I {Q^a}_I=0$, which causes
$\gamma_E$ to drop out of \eqref{eq:arat}-\eqref{eq:crat}.
Thus, all coefficients in \eqref{eq:arat}-\eqref{eq:crat} are manifestly rational.

In summary, we have
\begin{equation}
    c^a_{\Vec{n}}=\mathcal{A}^a_{\Vec{n}}c_{\Vec{n}}\, ,\quad \hat{c}^{ab}_{\Vec{n}}=\left(\mathcal{A}_{\vec{n}}^a \mathcal{A}_{\vec{n}}^b +\hat{\mathcal{B}}^{ab}_{\Vec{n}}\right)c_{\vec{n}}\, ,
\end{equation}
if all ${Q^a}_I n_a$ are non-negative.

For $\Vec{n}$ such that $k_I={Q^a}_I n_a<0$ for precisely one $I=I_0$, we reproduce from \cite{HKTY} that $c_{\Vec{n}}=0$, and
\begin{align}
    c^a_{\Vec{n}}&=-{Q^a}_{I_0} (-1)^{k_{I_0}} (|k_{I_0}|-1)! \times \frac{k_0!}{\prod_{I\neq I_0} k_I!}\, ,\\
    c^{ab}_{\Vec{n}}&=\hat{c}^{ab}_{\Vec{n}}=-\left({Q^a}_{I_0}\mathcal{A}_{\Vec{n},I_0}^b+ {Q^b}_{I_0}\mathcal{A}_{\Vec{n},I_0}^a\right) (-1)^{k_{I_0}} (|k_{I_0}|-1)! \times \frac{k_0!}{\prod_{I\neq I_0} k_I!}\, ,
\end{align}
in terms of
\begin{equation}
    \mathcal{A}_{\Vec{n},I_0}^a:={Q^a}_0 H_{k_0}-\sum_{I\neq I_0} {Q^a}_I H_{k_I}-{Q^a}_{I_0}H_{|k_{I_0}|-1}\, .
\end{equation}
Finally, for $k_I={Q^a}_I n_a<0$ for precisely two $I\in \{I_0,I_1\}$ we have $c_{\Vec{n}}=c^a_{\Vec{n}}=0$, and
\begin{equation}
    c^{ab}_{\Vec{n}}=\left({Q^a}_{I_0}{Q^b}_{I_1}+{Q^a}_{I_1}{Q^b}_{I_0}\right) (-1)^{k_{I_0}+k_{I_1}} (|k_{I_0}|-1)!  (|k_{I_1}|-1)!\times \frac{k_0!}{\prod_{I\notin \{I_0,I_1\}} k_I!}\, ,
\end{equation}
and $\hat{c}^{ab}_{\Vec{n}}\equiv c^{ab}_{\Vec{n}}$. In all other cases, i.e.~whenever ${Q^a}_I n_a<0$ for more than two values of $I$, one has $c_{\Vec{n}}=c^a_{\Vec{n}}=c^{ab}_{\Vec{n}}=\hat{c}^{ab}_{\Vec{n}}=0$.

\bibliographystyle{JHEPmod}
\bibliography{biblio}

\end{document}